\documentclass{ieeeoj}

\usepackage{packages}
\begin{document}
\receiveddate{XX Month, XXXX}
\reviseddate{XX Month, XXXX}
\accepteddate{XX Month, XXXX}
\publisheddate{XX Month, XXXX}
\currentdate{XX Month, XXXX}
\doiinfo{OJAP.2020.1234567}
\DeclareRobustCommand*{\IEEEauthorrefmark}[1]{%
  \raisebox{0pt}[0pt][0pt]{\textsuperscript{\footnotesize\ensuremath{#1}}}}

\title{Inverse Design and Experimental Verification of a Bianisotropic Metasurface Using Optimization and Machine Learning} % Somehow fit SWs into the title...

\author{Stewart Pearson\IEEEauthorrefmark{1}, Graduate Student Member, IEEE, Parinaz Naseri\IEEEauthorrefmark{1}, Graduate Student Member, IEEE, AND Sean V. Hum\IEEEauthorrefmark{1}, Senior MEMBER, IEEE}
\affil{The Edward S. Rogers Sr. Department of Electrical \& Computer Engineering, University of Toronto, Toronto, ON, Canada, M5S 3G8}

\corresp{CORRESPONDING AUTHOR: Stewart Pearson (e-mail: stewart.pearson@mail.utoronto.ca).}
\authornote{IEEE OJAP encourages responsible authorship practices and the provision of information about the specific contribution of each author\\[1em]
This work was supported by the Natural Sciences and Engineering Research Council (NSERC) of Canada.}
\markboth{Inverse Design and Experimental Verification of a Bianisotropic Metasurface Using Optimization and Machine Learning}{S. Pearson \textit{et al.}}

\begin{abstract}
Electromagnetic metasurfaces have attracted significant interest recently due to their low profile and advantageous applications. Practically, many metasurface designs start with a set of constraints for the radiated far-field, such as main-beam direction(s) and side lobe levels, and end with a non-uniform physical structure for the surface. This problem is quite challenging, since the required tangential field transformations are not completely known when only constraints are placed on the scattered fields. Hence, the required surface properties cannot be solved for analytically. Moreover, the translation of the desired surface properties to the physical unit cells can be time-consuming and difficult, as it is often a one-to-many mapping in a  large solution space. Here, we divide the inverse design process into two steps: a macroscopic and microscopic design step. In the former, we use an iterative optimization process to find the surface properties that radiate a far-field pattern that complies with specified constraints. This iterative process exploits non-radiating currents to ensure a passive and lossless design. In the microscopic step, these optimized surface properties are realized with physical unit cells using machine learning surrogate models. The effectiveness of this end-to-end synthesis process is demonstrated through measurement results of a beam-splitting prototype.
\end{abstract}



\maketitle

\section{INTRODUCTION}

\IEEEPARstart{E}{lectromagnetic} metasurfaces (EMMSs) are electrically-thin surfaces that enable exotic field transformations \cite{Epstein2016}. These field transformations are accomplished by manipulating the generalized sheet transition conditions (GSTCs), which relate the fields on one side of the surface to the other through the surface parameters \cite{Kuester2003}. Bianisotropic magneto-electric and electro-magnetic coupling terms can also be leveraged as a degree of freedom to implement any desired power-conserving field transformation. The surface parameters are typically realized by sub-wavelength meta-atoms composed of dielectric and/or conductor scatterers, fabricated with printed circuit board (PCB) technology. 

The design of an EMMS can be separated into two broad stages: macroscopic and microscopic design steps. Macroscopic design encompasses choosing appropriate surface parameters to realize the desired field transformation. Microscopic design involves realizing these spatially varying surface parameters using meta-atoms with specifically patterned scatterers. Unfortunately, the design processes for both the macroscopic and microscopic design stages are \textit{ad hoc}, time consuming, and cumbersome, as they generally rely on iterative simulations. As a result, there has been some research into both the macroscopic and microscopic design processes in order to improve them. 

% Macro recent developments from Budhu, Vasilieos, Brown add Bodehou maybe as well 
There have been many recent efforts to perform macroscopic optimization in a more systematic way. One approach uses an electromagnetic inversion algorithm to solve for the required electric and magnetic surface current densities to produce a desired far-field pattern \cite{Brown2020a, Narendra2021}. This method utilizes the equivalence principle, whereby enforcing Love's condition and local power conservation using a gradient-based method, a passive and lossless EMMS is derived. Upon completion, one arrives at a 3-layer surface admittance profile for an EMMS yielding a far-field pattern with good agreement with a target pattern. One drawback of this method is that it requires that the target far-field pattern be completely defined so that the pattern from the EMMS can be matched to it, rather than imposing application-specific constraints on the far-field. Budhu \emph{et al}. utilize the method of moments (MoM) to model multi-layer EMMSs and subsequently optimize them to achieve desired, fully-specified far-field patterns \cite{Budhu2022, Budhu2021a}. This was further extended to dual-band stacked metasurfaces \cite{Budhu2022}. Similar to the previously described method, this also requires fully defined fields rather than providing the ability to satisfy far-field constraints with a passive and lossless EMMS. In contrast to the two previous approaches, Ataloglou \emph{et al}. explicitly invoke auxiliary fields (AFs) to realize Taylor \cite{Ataloglou2021a} and uniform \cite{Ataloglou2021b} aperture distributions. More recently, this method was used to optimize a MoM-based meta-wire structure with an integrated feed for far-field beam forming and MIMO applications \cite{Xu2021}. This approach optimizes the AFs by attempting to satisfy a fully specified desired aperture field distribution. The approach requires a fully defined target aperture field and so cannot optimize for loosely defined objectives in its current form. All of these works in their present form require a fully formed target far-field pattern or aperture field. This is slightly disconnected from typical antenna design applications where the designer is given far-field specifications, rather than an exact pattern, to match. 

% Micro recent developments
Microscopic optimization conventionally involves choosing a certain meta-atom design with adequate degrees of freedom to arbitrarily manipulate the amplitude and phase of incoming electric field based on its polarization. The meta-atom selection is typically accomplished by conducting iterative simulations of different structures based on empirical methods to converge to a successful candidate. Moreover, the properties of the meta-atoms are derived assuming local periodicity. Therefore, in a quasi-periodic EMMS, to reduce mutual coupling between adjacent meta-atoms, usually one type of structure is selected and the dimension of its scatterers are tuned to realize the desired scattering properties. Non-local design procedures are much less common because they require accurate models of the EMMS as a whole. However, recently, there has been some work to streamline this process.

Inspired by the revolution that data-driven machine learning methods have made in material informatics applied to fields such as quantum materials, pharmaceuticals, and chemistry \cite{GomezBombarelli2018}, deep  machine learning \cite{Prado2018a,Qiu2019,Richard2017,Kampouridou2020,Caputo2009,Robustillo2012} and statistical learning \cite{Salucci2018a} methods can help to build surrogate models that can provide fast predictions of the properties of each unit cell. Moreover, several machine learning techniques have been proposed to tackle the challenges of the microscopic design step \cite{Liu2018,Shi2020,Jiang2019,Jiang2019a,An2021,Ma2019,Gosal2015,Gosal2016,Oliveri2020,Yeung2021}. Some of these proposed methods deal with the inverse design of uniform EMMSs, where the impact of inter-cell mutual coupling is less of an issue \cite{Liu2018,Shi2020,An2021,Ma2019,Naseri2021,Hodge2019,Hodge2021}, while the rest optimize over a simple solution space that is composed of only one scatterer shape \cite{Gosal2015, Gosal2016,Oliveri2020,Yeung2021}. It is worth noting that dielectric optical EMMSs can be designed using global optimization methods due to the analytical relation between the scatterers' properties and the EMMS's scattering parameters \cite{Jiang2019}. However, due to the lack of such relations in EMMSs composed of metallic scatterers, the inverse design of a heterogeneous bianisotropic EMMS is more challenging. Nonetheless, solving this problem more efficiently has led to the proposal of systematic approaches that provide both the optimized surface properties and the actual physical unit cells for specific applications \cite{Budhu2022,Pfeiffer2014,Oliveri2020}.

% Explain how conventional designs require ad hoc methods for both the macro and micro sides. Explain lack of end-to-end designs
In order to make the EMMS design process more streamlined, we aim to use the integrated macroscopic and microscopic optimizers the authors previously presented \cite{Naseri2021a} to solve for an EMMS that forms two beams. Although the results were previously verified in simulation using Ansys HFSS, we will go one step further to experimentally verify them using a fabricated EMMS illuminated by a standard gain horn (SGH) in a near-field chamber. Importantly, we will also investigate how the macroscopic optimizer utilizes AFs as an extra degree of freedom in its solution.

% Talk about the structure of the paper, macro, AFs, micro, experimental results, conclusions 
We will begin by giving a brief overview of the alternating direction method of multipliers (ADMM)-based macroscopic optimizer in \Cref{sec:macro_opt}. Following this, we will show how the macroscopic optimizer leverages AFs to ensure a passive and lossless design \Cref{sec:LeveragingSWs}. We will then outline how the microscopic optimizer uses deep-learning neural networks and particle swarm optimization to solve for unit cells in \Cref{sec:MicroOverview}. We next utilize the macroscopic and microscopic optimization steps to synthesize a two-beam EMMS in \Cref{Sec:Example}. Experimental verification of the optimized two-beam EMMS is shown in \Cref{sec:ExperimentalVerif}. Lastly, some concluding remarks are offered in \Cref{sec:Conclusion}.

\section{ADMM-BASED MACROSCOPIC OPTIMIZER}\label{sec:macro_opt}
% The usual stuff about how we use ADMM to solve the problem and some details on the MoM model. Explain the 2D setup as per usual (1 page)
The macroscopic optimizer is similar to the ones reported previously \cite{Pearson2021a, Naseri2021a}. It is formulated using a homogenized two-dimensional EMMS model constructed with the MoM. This homogenized model captures omega-type bianisotropic behavior of an EMMS. Importantly, the model incorporates the important mutual coupling and edge effects for EMMSs. 

% Details on macro model
The macroscopic model is constructed using the two-dimensional MoM. We first describe the electric and magnetic surface current densities on the EMMS using $N$ pulse basis functions yielding the expansion coefficients $\textit{\textbf{I}}^e \in \mathbb{C}^N$ and $\textit{\textbf{I}}^m \in \mathbb{C}^N$ respectively. We then use point matching, in conjunction with the incident field across the EMMS, to construct the MoM coupling matrices $[{\textbf{Z}}^e] \in \mathbb{C}^{N\times N}$ and $[{\textbf{Z}}^m] \in \mathbb{C}^{N\times N}$ respectively. These matrices relate the incident field with the induced surface current densities.

\begin{figure}[!t]
    \centering
    \resizebox{0.3\textwidth}{!}{%
        \begin{tikzpicture}
	\begin{pgfonlayer}{nodelayer}
		\node [style=none] (0) at (-0.25, 8) {};
		\node [style=none] (1) at (0.25, 8) {};
		\node [style=none] (2) at (-0.25, -8) {};
		\node [style=none] (3) at (0.25, -8) {};
		\node [style=none] (4) at (0, 0) {};
		\node [style=none] (5) at (0, 9) {};
		\node [style=none] (6) at (5, 0) {};
		\node [style=none] (7) at (-5, 6) {};
		\node [style=none] (8) at (-5, 4) {};
		\node [style=none] (9) at (-5, 2) {};
		\node [style=none] (10) at (-5, 0) {};
		\node [style=none] (11) at (-5, -2) {};
		\node [style=none] (12) at (-5, -4) {};
		\node [style=none] (13) at (-5, -6) {};
		\node [style=none] (14) at (-4, 6) {};
		\node [style=none] (15) at (-4, 4) {};
		\node [style=none] (16) at (-4, 2) {};
		\node [style=none] (17) at (-4, 0) {};
		\node [style=none] (18) at (-4, -2) {};
		\node [style=none] (19) at (-4, -4) {};
		\node [style=none] (20) at (-4, -6) {};
		\node [style=none] (21) at (-3, 6) {};
		\node [style=none] (22) at (-3, 4) {};
		\node [style=none] (23) at (-3, 2) {};
		\node [style=none] (24) at (-3, 0) {};
		\node [style=none] (25) at (-3, -2) {};
		\node [style=none] (26) at (-3, -4) {};
		\node [style=none] (27) at (-3, -6) {};
		\node [style=none] (28) at (-6, 4) {};
		\node [style=none] (29) at (-2, 4) {};
		\node [style=none] (30) at (-6, 0) {};
		\node [style=none] (31) at (-2, 0) {};
		\node [style=none] (32) at (-5.75, -4) {};
		\node [style=none] (33) at (-1.75, -4) {};
		\node [style=none] (34) at (-4, 7) {};
		\node [style=none] (35) at (-4, 7) {\Huge Incident Wave};
		\node [style=none] (36) at (0, 9.5) {\Huge$y$};
		\node [style=none] (37) at (5.25, 0) {\Huge$z$};
		\node [style=none] (38) at (-0.75, 8) {};
		\node [style=none] (39) at (0.75, 8) {};
		\node [style=none] (40) at (1.5, 8) {\Huge $\;W/2$};
		\node [style=none] (41) at (1.5, -8) {\Huge $\;-W/2$};
		\node [style=none] (42) at (0.75, -8) {};
		\node [style=none] (43) at (-0.75, -8) {};
		\node [style=none] (44) at (3, 4) {};
		\node [style=none] (45) at (1.25, 1.5) {};
		\node [style=none] (46) at (1.75, 0) {};
		\node [style=none] (48) at (1, 0.5) {\Huge $\theta$};
	\end{pgfonlayer}
	\begin{pgfonlayer}{edgelayer}
		\draw [style={EMMS_edge}] (3.center)
			 to (2.center)
			 to (0.center)
			 to (1.center)
			 to cycle;
		\draw [style=axis] (4.center) to (5.center);
		\draw [style=axis] (4.center) to (6.center);
		\draw [style={wave_edge}] (7.center)
			 to [bend left] (8.center)
			 to [bend right] (9.center)
			 to [bend left] (10.center)
			 to [bend right] (11.center)
			 to [bend left] (12.center)
			 to [bend right, looseness=0.75] (13.center);
		\draw [style={wave_edge}] (14.center)
			 to [bend left] (15.center)
			 to [bend right] (16.center)
			 to [bend left] (17.center)
			 to [bend right] (18.center)
			 to [bend left] (19.center)
			 to [bend right, looseness=0.75] (20.center);
		\draw [style={wave_edge}, bend left] (21.center) to (22.center);
		\draw [style={wave_edge}, bend right] (22.center) to (23.center);
		\draw [style={wave_edge}, in=60, out=-60] (23.center) to (24.center);
		\draw [style={wave_edge}, bend right] (24.center) to (25.center);
		\draw [style={wave_edge}, bend left] (25.center) to (26.center);
		\draw [style={wave_edge}, bend right, looseness=0.75] (26.center) to (27.center);
		\draw [style={wave_direction_edge}] (28.center) to (29.center);
		\draw [style={wave_direction_edge}] (30.center) to (31.center);
		\draw [style={wave_direction_edge}] (32.center) to (33.center);
		\draw [style={meas_edge}] (38.center) to (39.center);
		\draw (43.center) to (42.center);
		\draw [style={angle_edge}] (4.center) to (44.center);
		\draw [style={angle_edge_arrow}, bend left=60] (45.center) to (46.center);
	\end{pgfonlayer}
\end{tikzpicture}
    }%
    \caption{The EMMS is finite with width $W$ and coincident with the $y$-axis, while uniform and infinite in the $x$-direction.}
    \label{fig:EMMS_config}
  \end{figure}
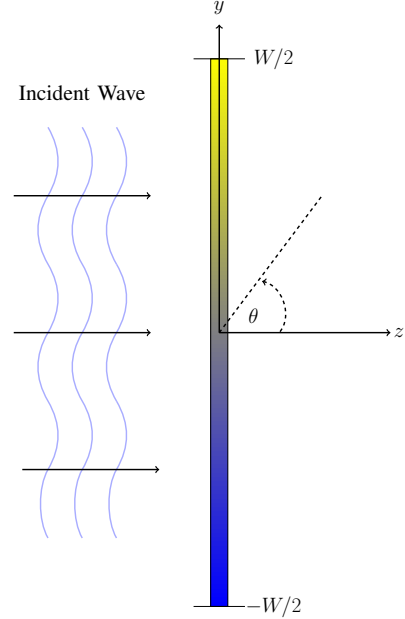

We consider a one-dimensional EMMS located along the $y$-axis, which is uniform in the $x$-direction. This configuration is shown in \Cref{fig:EMMS_config}. In this paper we will restrict our consideration to TE-polarized examples for a passive, lossless, omega-type bianisotropic EMMS. These surface parameters can be described in a few different ways, but we will use the surface electric impedance ($Z_{se}$), magnetic admittance ($Y_{sm}$), and electro-magnetic coupling ($K_{em}$) representation. For a passive and lossless structure, $\Re\{ Z_{se}\}=\Re\{Y_{sm}\}=\Im\{K_{em}\}=0$.  As a result, the surface parameters in the two-dimensional problem reduce to a scalars $jX_{se}$, $jB_{sm}$, and $K_{em}$. This yields the matrix equations \cite{Pearson2021a, Naseri2021a}
\begin{align}
\widetilde{\textit{\textbf{E}}}^{inc} &= [\widetilde{\textbf{Z}}_e]\widetilde{\textit{\textbf{I}}}^e+[\widetilde{\textbf{X}}_{se}]\widetilde{\textit{\textbf{I}}}^e+ [\widetilde{\textbf{K}}_{em}]\widetilde{\textit{\textbf{I}}}^m  \label{eq:MoME}\\
\widetilde{\textit{\textbf{H}}}^{inc} &= [\widetilde{\textbf{Z}}_m]\widetilde{\textit{\textbf{I}}}^m+  [\widetilde{\textbf{B}}_{sm}]\widetilde{\textit{\textbf{I}}}^m -[\widetilde{\textbf{K}}_{em}]\widetilde{\textit{\textbf{I}}}^e, \label{eq:MoMH}
\end{align}
where $\widetilde{\textit{\textbf{E}}}^{inc}$ and $\widetilde{\textit{\textbf{H}}}^{inc}$ are the incident electric and magnetic field, $[\widetilde{\textbf{Z}}_e]$ and $[\widetilde{\textbf{Z}}_m]$ are the two MoM matrices for calculating the scattered field, and $\widetilde{\textit{\textbf{I}}}^e$ and $\widetilde{\textit{\textbf{I}}}^m$ are the electric and magnetic surface current density coefficients. The $\widetilde{\cdot}$ notation denotes the separation of real and imaginary portions of a complex number \cite{Pearson2021a} as optimization with complex numbers adds another layer of complexity. There is no problem information lost when performing this conversion. 

Satisfaction of \eqref{eq:MoME} and \eqref{eq:MoMH} ensures that the EMMS is passive, lossless, and omega-type bianisotropic \emph{by construction}. Optimizing only the surface current densities for far-field objectives would likely result in an active and/or lossy solution for the surface parameters. As a result, we need to consider both the surface parameters and surface currents when optimizing to ensure that the solution satisfies \eqref{eq:MoME} and \eqref{eq:MoMH}.

In order to impose constraints on the far-field, we require matrices $[\widetilde{\textbf{G}}^e]$ and $[\widetilde{\textbf{G}}^e]$ to transform the surface current density coefficients to the far-field. As a result, our expression for the total field becomes
\begin{align}
\widetilde{\textbf{E}}_{ff}^{tot} = [\widetilde{\textbf{G}}^e]\widetilde{\textbf{\textit{I}}}^e+[\widetilde{\textbf{G}}^m]\widetilde{\textbf{\textit{I}}}^m + \widetilde{\textbf{E}}_{ff}^{inc}\label{eq:Eff}
\end{align}
where $\widetilde{\textbf{E}}_{ff}^{inc}$ is the far-field of the incident field's contribution across the extent of the EMMS.

From this, we can construct an optimization program

\begin{subequations} \label{eq:EMMS_total_form}
\begin{align}
\minimize_{\substack{\textit{\textbf{I}}^e,\textit{\textbf{I}}^m,\\ [\textbf{X}_{se}],[\textbf{B}_{sm}],[\textbf{K}_{em}],\\{\bm{\gamma}}_{D^e},{\bm{\gamma}}_{D^m}}} \begin{split}&\;\alpha_{MB}f_{MB}(MB)+\alpha_{NU}f_{NU}(NU)\\&\;+\alpha_{D}f_{D}\end{split}\\
\textrm{subject to}  \quad\;\;\;   & \; \widetilde{\textit{\textbf{E}}}^{inc} = [\widetilde{\textbf{Z}}_e]\widetilde{\textit{\textbf{I}}}^e+[\widetilde{\textbf{X}}_{se}]\widetilde{\textit{\textbf{I}}}^e + [\widetilde{\textbf{K}}_{em}]\widetilde{\textit{\textbf{I}}}^m \label{E_mom_eqn}\\
\begin{split}& \; \widetilde{\textit{\textbf{H}}}^{inc} = [\widetilde{\textbf{Z}}_m]\widetilde{\textit{\textbf{I}}}^m+   [\widetilde{\textbf{B}}_{sm}]\widetilde{\textit{\textbf{I}}}^m -[\widetilde{\textbf{K}}_{em}]\widetilde{\textit{\textbf{I}}}^e\end{split} \label{H_mom_eqn}\\
\begin{split}&\;\vert [\widetilde{\textbf{G}}^e](SL)\widetilde{\textbf{\textit{I}}}^e+[\widetilde{\textbf{G}}^m](SL)\widetilde{\textbf{\textit{I}}}^m \\ &\;+ \widetilde{\textbf{E}}_{ff}^{inc}(SL) \vert \leq \bm{\tau}\end{split} \label{eq:SLL_const} \\
& \; \vert [\textbf{D}]\widetilde{\textbf{\textit{I}}}^e\vert = \textbf{D}_{max}^e + \bm{\gamma}_{D^e} \label{eq:Ie_curv}\\
& \; \vert [\textbf{D}]\widetilde{\textbf{\textit{I}}}^m\vert = \textbf{D}_{max}^m + \bm{\gamma}_{D^m} \label{eq:Im_curv}\\
& \; \textbf{\textit{X}}_{se}^{min} \leq \textrm{diag(}[\textbf{X}_{se}]\textrm{)} \leq \textbf{\textit{X}}_{se}^{max}\label{eq:Xe_lim}\\
& \; \textbf{\textit{B}}_{sm}^{min} \leq \textrm{diag(}[\textbf{B}_{sm}]\textrm{)} \leq \textbf{\textit{B}}_{sm}^{max}\label{eq:Bm_lim}\\
& \;\textbf{\textit{K}}_{em}^{min} \leq \textrm{diag(}[\textbf{K}_{em}]\textrm{)} \leq \textbf{\textit{K}}_{em}^{max}\label{eq:Kem_lim}.
\end{align}
\end{subequations}

The optimization variables in this case are the surface electric and magnetic current densities ($\textit{\textbf{I}}^e,\textit{\textbf{I}}^m$), the passive and lossless surface parameters ($[\textbf{X}_{se}],[\textbf{B}_{sm}],[\textbf{K}_{em}]$), and the slack variables (${\bm{\gamma}}_{D^e},{\bm{\gamma}}_{D^m}$). The side lobe level $\tau$, over specified angles $SL$, is constrained in \eqref{eq:SLL_const}. The second derivative of the surface current densities are constrained with \eqref{eq:Ie_curv} and \eqref{eq:Im_curv} in order to aid numerical stability. The limits of the achievable surface parameters are also constrained in \eqref{eq:Xe_lim}, \eqref{eq:Bm_lim}, and \eqref{eq:Kem_lim} with predefined limits $\textbf{\textit{X}}_{se}^{min}$, $\textbf{\textit{X}}_{se}^{max}$, $\textbf{\textit{B}}_{sm}^{min}$, $\textbf{\textit{B}}_{sm}^{max}$, $\textbf{\textit{K}}_{em}^{min}$, and $\textbf{\textit{K}}_{em}^{max}$. The terms $\alpha_{MB}$, $\alpha_{NU}$, and $\alpha_{D}$ are the hyper-parameters, which can be used to weight the relative values of the components of the objective function to aid convergence. The function 
\begin{align}
\begin{split}
    f_{MB}(MB) =& \Vert [\widetilde{\textbf{G}}^e](MB)\widetilde{\textbf{\textit{I}}}^e+[\widetilde{\textbf{G}}^m](MB)\widetilde{\textbf{\textit{I}}}^m \\& + \widetilde{\textbf{E}}_{ff}^{inc}(MB) - MB_{level}\Vert_2^2
\end{split}
\end{align}
aims to match the field level at an an angle $MB$ to a desired field level $MB_{level}$. Similarly, the function
\begin{align}
\begin{split}
    f_{NU}(NU) =& \Vert [\widetilde{\textbf{G}}^e](NU)\widetilde{\textbf{\textit{I}}}^e+[\widetilde{\textbf{G}}^m](NU)\widetilde{\textbf{\textit{I}}}^m \\& + \widetilde{\textbf{E}}_{ff}^{inc}(NU) \Vert_2^2
\end{split}
\end{align} 
simply aims to minimize the field at an angle $NU$, thereby producing a null in the far-field pattern there. Lastly, the function
\begin{align}
    f_D = \Vert (\gamma_{D^e})_+\Vert_2^2+\Vert (\gamma_{D^m})_+\Vert_2^2,
\end{align}
where $(\cdot)_+$ returns the value of the input if it is positive and zero for negative inputs, minimizes the positive values of the slack variables $\gamma_{D^e}$ and $\gamma_{D^m}$. As a result, this function aims to minimize the amount of curvature in the current densities that exceeds the user-defined values $\textbf{D}_{max}^e$ and $\textbf{D}_{max}^m$. Using slack variables for \eqref{eq:Ie_curv} and \eqref{eq:Im_curv} allows for some violation of the constraint during optimization to aid convergence. 

This model contains non-convex bilinear terms in \eqref{E_mom_eqn} and \eqref{H_mom_eqn}, which cannot be optimized with fast and efficient convex solvers. To circumvent this, we use a non-linear optimization technique called the alternating direction method of multipliers (ADMM). The algorithm first forms the augmented Lagrangian, transferring the equality constraints in \eqref{E_mom_eqn} and \eqref{H_mom_eqn} to the objective. The two equality constraints are combined into a single function $f_Z$ and added to the objective as
\begin{align}
\begin{split}
&f_{Z_e} = \frac{\rho}{2}\Vert -\widetilde{\textit{\textbf{E}}}^{inc} + [\widetilde{\textbf{Z}}_e]\widetilde{\textit{\textbf{I}}}^e + [\widetilde{\textbf{X}}_{se}]\widetilde{\textit{\textbf{I}}}^e + [\widetilde{\textbf{K}}_{em}]\widetilde{\textit{\textbf{I}}}^m  \Vert_2^2 +\\
&\quad\quad \bm{\mu}_{Z_e}^T\left( -\widetilde{\textit{\textbf{E}}}^{inc} + [\widetilde{\textbf{Z}}_e]\widetilde{\textit{\textbf{I}}}^e + [\widetilde{\textbf{X}}_{se}]\widetilde{\textit{\textbf{I}}}^e + [\widetilde{\textbf{K}}_{em}]\widetilde{\textit{\textbf{I}}}^m \right)\\
\end{split}\\
\begin{split}
&f_{Z_m} = \frac{\rho}{2}\Vert -\widetilde{\textit{\textbf{H}}}^{inc} + [\widetilde{\textbf{Z}}_m]\widetilde{\textit{\textbf{I}}}^m+   [\widetilde{\textbf{B}}_{sm}]\widetilde{\textit{\textbf{I}}}^m -[\widetilde{\textbf{K}}_{em}]\widetilde{\textit{\textbf{I}}}^e \Vert_2^2 +\\
&\quad\quad \bm{\mu}_{Z_m}^T\left( -\widetilde{\textit{\textbf{H}}}^{inc} + [\widetilde{\textbf{Z}}_m]\widetilde{\textit{\textbf{I}}}^m+   [\widetilde{\textbf{B}}_{sm}]\widetilde{\textit{\textbf{I}}}^m -[\widetilde{\textbf{K}}_{em}]\widetilde{\textit{\textbf{I}}}^e \right)
\end{split}\\
&f_z = f_{Z_e} + \beta f_{Z_m}
\end{align}
where $f_Z$ is a weighted sum of the two functions $f_{Z_e}$ and $f_{Z_m}$, $\rho$ is a user-defined penalty parameter, $\beta$ is a real number used to scale $f_{Z_m}$, and $\bm{\mu}_{Z_e}$ and $\bm{\mu}_{Z_m}$ are dual variables. The dual variables are updated each iteration $i$ according to 
\begin{align}
\bm{\mu}_{Z_e}^{i+1} &= \bm{\mu}_{Z_e}^{i}+ \rho\left( -\widetilde{\textit{\textbf{E}}}^{inc} + [\widetilde{\textbf{Z}}_e]\widetilde{\textit{\textbf{I}}}^e + [\widetilde{\textbf{X}}_{se}]\widetilde{\textit{\textbf{I}}}^e + [\widetilde{\textbf{K}}_{em}]\widetilde{\textit{\textbf{I}}}^m  \right)\\
\bm{\mu}_{Z_m}^{i+1} &= \bm{\mu}_{Z_m}^{i}+ \rho\left( -\widetilde{\textit{\textbf{H}}}^{inc} + [\widetilde{\textbf{Z}}_m]\widetilde{\textit{\textbf{I}}}^m+   [\widetilde{\textbf{B}}_{sm}]\widetilde{\textit{\textbf{I}}}^m -[\widetilde{\textbf{K}}_{em}]\widetilde{\textit{\textbf{I}}}^e \right).
\end{align}
This new optimization program has the form
\begin{subequations} \label{eq:EMMS_total_form_cvx}
\begin{align}
\minimize_{\substack{\textit{\textbf{I}}^e,\textit{\textbf{I}}^m,\\ [\textbf{X}_{se}],[\textbf{B}_{sm}],[\textbf{K}_{em}],\\{\bm{\gamma}}_{D^e},{\bm{\gamma}}_{D^m}}} \begin{split}&\;\alpha_{MB}f_{MB}(MB)+\alpha_{NU}f_{NU}(NU)\\&\;+\alpha_{D}f_{D}+f_Z\end{split} \\
\textrm{subject to}  \quad\;\;\;   
\begin{split}&\;\vert [\widetilde{\textbf{G}}^e](SL)\widetilde{\textbf{\textit{I}}}^e+[\widetilde{\textbf{G}}^m](SL)\widetilde{\textbf{\textit{I}}}^m \\ &\;+ \widetilde{\textbf{E}}_{ff}^{inc}(SL) \vert \leq \bm{\tau}\end{split}  \\
& \; \vert [\textbf{D}]\widetilde{\textbf{\textit{I}}}^e\vert = \textbf{D}_{max}^e + \bm{\gamma}_{D^e} \\
& \; \vert [\textbf{D}]\widetilde{\textbf{\textit{I}}}^m\vert = \textbf{D}_{max}^m + \bm{\gamma}_{D^m}\\
& \; \textbf{\textit{X}}_{se}^{min} \leq \textrm{diag(}[\textbf{X}_{se}]\textrm{)} \leq \textbf{\textit{X}}_{se}^{max}\\
& \; \textbf{\textit{B}}_{sm}^{min} \leq \textrm{diag(}[\textbf{B}_{sm}]\textrm{)} \leq \textbf{\textit{B}}_{sm}^{max}\\
& \;\textbf{\textit{K}}_{em}^{min} \leq \textrm{diag(}[\textbf{K}_{em}]\textrm{)} \leq \textbf{\textit{K}}_{em}^{max}.
\end{align}
\end{subequations}

ADMM optimizes \eqref{eq:EMMS_total_form_cvx} for the surface electric and magnetic current densities and then surface parameters with each iteration. This has the effect of breaking the original non-convex problem into smaller, convex ones. Solutions which satisfy the desired problem specifications can usually be found in roughly one hundred iterations.
% TODO: Show the spectrum of AFs from this retro-reflection example to illustrate how we use AFs
% You could also try showing real ZY, imag K resulting from using visible currents only

\section{LEVERAGING AUXILIARY FIELDS TO ENSURE SATISFACTIONS OF GSTCS} \label{sec:LeveragingSWs}
In previous work by Epstein \textit{et al.} \cite{Epstein2016b}, it is shown that leveraging the AFs introduced on either side of a metasurface can help satisfy local power conservation. This allows for arbitrary field transformations with passive and lossless omega-type bianisotropic surfaces. These AFs are evanescent and as a result, do not contribute to the far-field pattern of interest. Recently, work has been done to optimize these AFs to tailor aperture fields for beamforming \cite{Ataloglou2021a} along with Chebyshev and Taylor array patterns \cite{Ataloglou2020a}. In contrast, the macroscopic optimizer outlined in \Cref{sec:macro_opt} does not directly optimize the AFs to achieve these transformations with a passive and lossless omega-type bianisotropic surface. The macroscopic optimizer instead indirectly leverages AFs to satisfy the passive and lossless GSTCs equality constraints through jointly optimizing the surface currents and parameters. To show that the optimizer is indeed leveraging AFs, we will first decompose the surface current densities into their radiating and the non-radiating components used to satisfy the GSTCs \cite{Salucci2018}. We will then show an example of extreme angle reflection and compare it to an analytically derived solution, which requires AFs \cite{Epstein2016b}.
% some details on proving AFs exist using spectrum of currents and/or presence of invisible currents. Show an example of how the GSTCs are violated when you remove invisible currents. Show an example of the optimizer performing retro-reflection or something. Maybe even pattern matching what Vasilieos did or make up my own currents to match... (1-2 pages)

\subsection{SEPARATING RADIATING AND NON-RADIATING CURRENT CONTRIBUTIONS}
% Show removing the invisible part leaves the radiated far field intact but messes up the equality constraints requiring active/lossy behaviour
The surface current densities contain both a radiating component, which is detectable in the far-field, and a non-radiating component, which is evanescent. Salucci \textit{et al.} \cite{Salucci2018} illustrate that singular value decomposition (SVD) can be performed on the far-field transformation matrices $[\textbf{G}^e]$ and $[\textbf{G}^m]$, shown in \eqref{eq:Eff}, to find the most significant current components contributing to the far-field pattern. To perform this, we calculate the scattered field from surface electric and magnetic current densities and then perform SVD as
\begin{subequations}
\begin{align}
    \textbf{\textit{E}}_{ff}^e +\textbf{\textit{E}}_{ff}^m &= \begin{bmatrix}\textbf{G}^e \;\textbf{G}^m \end{bmatrix} \begin{bmatrix}\textbf{\textit{I}}^e \\ \textbf{\textit{I}}^m, \end{bmatrix}\\
    \textbf{\textit{E}}_{ff}^{scat} &= \textbf{G} \textbf{\textit{I}},\\
    \textbf{\textit{E}}_{ff}^{scat} &= \textbf{U} \bm{\Sigma} \textbf{V}^H \textbf{\textit{I}},
\end{align}
\end{subequations}
where $^H$ is the complex conjugate transpose. If we take the most significant singular values, defined by a variable $\xi$, contributing to this radiated field, we can find the radiating surface current densities ($\textbf{\textit{I}}_{rad}$) as
\begin{align}
    \textbf{\textit{I}}_{rad} = \textbf{V}_\xi \bm{\Sigma}^{-1}_\xi \textbf{U}^H_\xi \textbf{\textit{E}}_{ff}^{scat}
\end{align}

These components are termed ``radiating", while the remaining currents (i.e. the currents corresponding to the less significant singular values) are ``non-radiating". The $\cdot_\xi$ notation signifies the truncated SVD matrices. These non-radiating currents can be associated with the evanescent AFs. When removed, the far-field radiation pattern remains essentially unchanged. While Salucci \textit{et al.} use non-radiating currents for various purposes, e.g. to impose forbidden regions for currents, we use them as degrees of freedom to achieve the required passivity and losslessness imposed by \eqref{eq:MoME} and \eqref{eq:MoMH}. Indeed, it is shown that these additional degrees of freedom are required for some wave transformations, such as perfect reflection and transmission, which redistribute power across the EMMS \cite{Epstein2016b}.

\subsection{EXAMPLE OF EXTREME ANGLE REFLECTION EMMS REQUIRING AUXILIARY FIELDS}
Epstein \textit{et al.} \cite{Epstein2016b} demonstrate that AFs are required to perform perfect reflection from an incident plane wave on an infinite metasurface. Our MoM model is finite, which is not accounted for in the example by Epstein \textit{et al}. Nevertheless, introducing AFs significantly improves the performance. Here we aim to achieve similar performance to their perfect reflection far-field pattern using the macroscopic optimizer to show that the optimizer leverages AFs. We will then evaluate the near field spectrum and use SVD to verify the extent to which AFs are used to satisfy the GSTCs. We attempt perfect reflection to $\theta=108^\circ$ from an incident plane wave at $\theta = 0^\circ$ for the EMMS shown in \Cref{fig:EMMS_config}. 

\subsubsection{PROBLEM FORMULATION}\label{sec:Retro_ProbForm}
In order to perform extreme angle reflection, we must supply a set of far-field goals for the optimizer described in \Cref{sec:macro_opt} to attempt to satisfy. Because our task is simply to form a beam at $\theta = 108^\circ$, our only far-field criteria will be to match the total radiated field to a predetermined value (${MB}_{level}$) at that angle. In order to maintain numerical stability, we will also impose constraints on the maximum allowable curvature of the electric ($\textbf{D}^e_{max}$) and magnetic surface current densities ($\textbf{D}^m_{max}$). The full list of optimization parameters is shown in \Cref{table:XtremeRefl_Crit}.

\begin{table}[!t]
\caption{Extreme-angle reflection optimization parameters}
\label{table:XtremeRefl_Crit}
\centering
\begin{tabular}{|c||c|}
\hline
 & \bfseries Value \\
\hline
\bfseries Surface Width ($W$) & $6\lambda_0$ \\
\hline
\bfseries Incident Field  & $e^{-jkz} \hat{x}$ [V/m] \\
\bfseries ($E^{inc}$) & uniform along the EMMS \\
\hline
\bfseries Angular Sampling Points ($M$) & $361$ \\
\hline
\bfseries Spatial Sampling Points ($N$) & $300$ \\
\hline
\bfseries Max Iterations & $100$ \\
\hline
\bfseries Initial & $ ([\widetilde{\textbf{X}}_{se}]^0,[\widetilde{\textbf{B}}_{sm}]^0,[\widetilde{\textbf{K}}_{em}]^0) = \textbf{0}, $\\
\bfseries Conditions & $\rho = 10, \beta = 1000, (\bm{\mu}_{\textbf{Z}_e}^0, \bm{\mu}_{\textbf{Z}_m}^0) = \textbf{0}$ \\
\hline
\bfseries $\left\lbrace \alpha_{MB},\alpha_{NU},\alpha_{D^E},\alpha_{D^H}\right\rbrace$ & $\left\lbrace 150,0,5,1\right\rbrace$ \\
\hline
\bfseries Main Lobe Angle  ($MB$) & $\theta = 108^{\circ}$ \\
\hline
\bfseries Main Lobe Level ($MB_{level}$) & 6 [V/m] \\
\hline
\bfseries $\left\lbrace \textbf{D}^e_{max},\textbf{D}^m_{max} \right\rbrace$ & $\left\lbrace \textbf{0.5},\textbf{225}\right\rbrace$ \\
\hline
\end{tabular}
\end{table}

\subsubsection{RESULTS}
\begin{figure}[!t]
\centering
\includegraphics[width=0.5\textwidth]{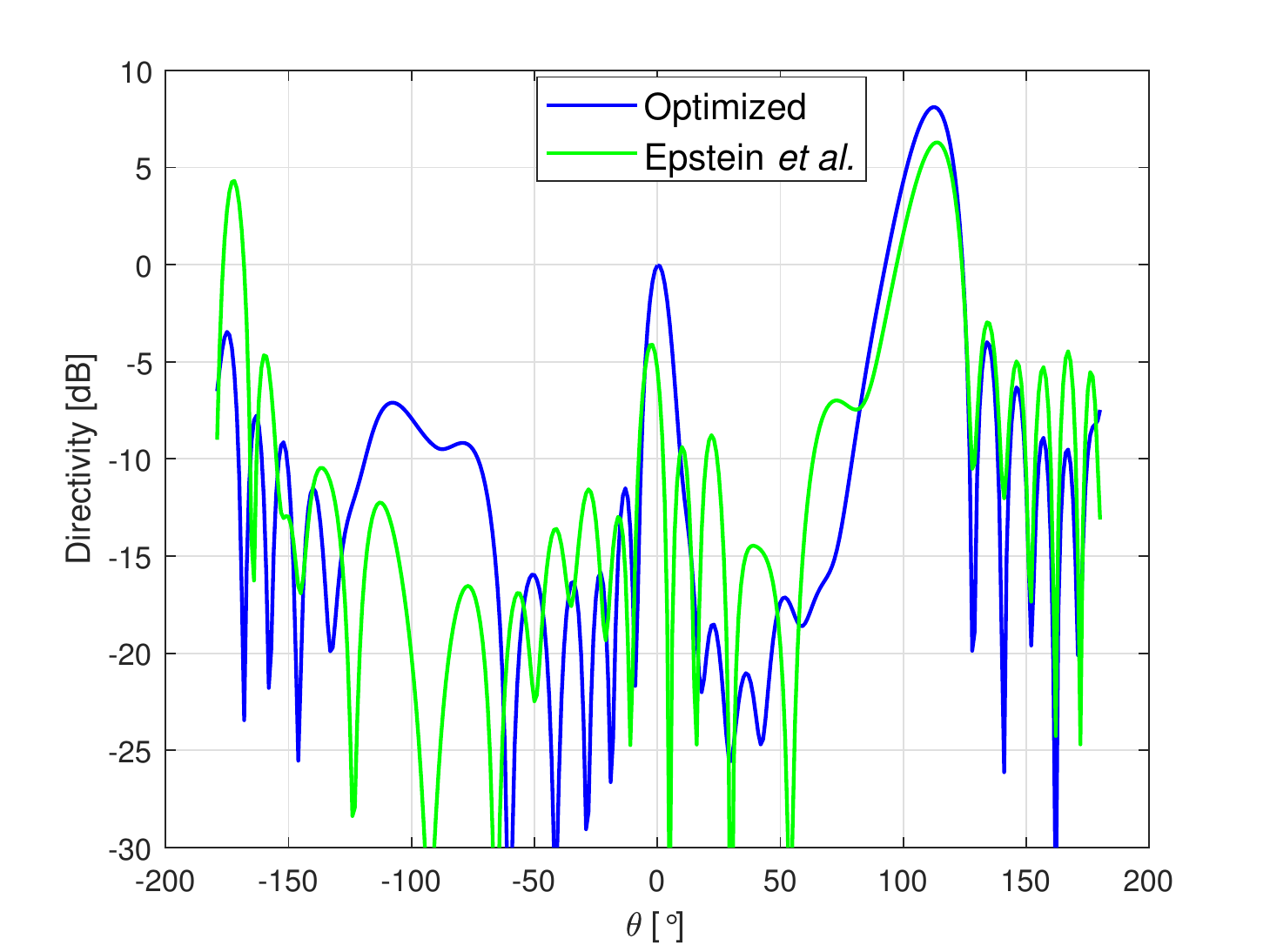}
\caption{Directivity of extreme angle reflection towards $\theta=108^\circ$ from an incident uniform illumination at $\theta = 0^\circ$. The optimized results are from the ADMM-based optimizer and analytical results are from Epstein \textit{et al.} \cite{Epstein2016b} using AFs.}
\label{fig:XTremeAngle_Directivity}
\end{figure}

After optimizing the surface parameters with the program described in \Cref{sec:Retro_ProbForm}, we arrive at the far-field pattern shown in \Cref{fig:XTremeAngle_Directivity}. We have also shown in the far-field results of an analytical formulation derived by Epstein \textit{et al.} \cite{Epstein2016b}, which uses AFs to perform perfect extreme angle reflection for an infinite surface. 

The analytical formulation will not exhibit perfect reflection because the surface parameters were derived assuming an infinite surface, while the surface considered is only $6\lambda_0$ long. This is evident from the spurious lobes apparent in the far-field directivity plot in \Cref{fig:XTremeAngle_Directivity}. The optimized far-field has slightly greater directivity compared to the analytic formulation. The improved directivity suggests that the optimizer factors in the finite width of the surface and adjusts the generation of the AFs to satisfy the far-field goals. 

\begin{figure}
     \centering
     \begin{subfigure}[b]{0.45\textwidth}
         \centering
         \includegraphics[width=\textwidth]{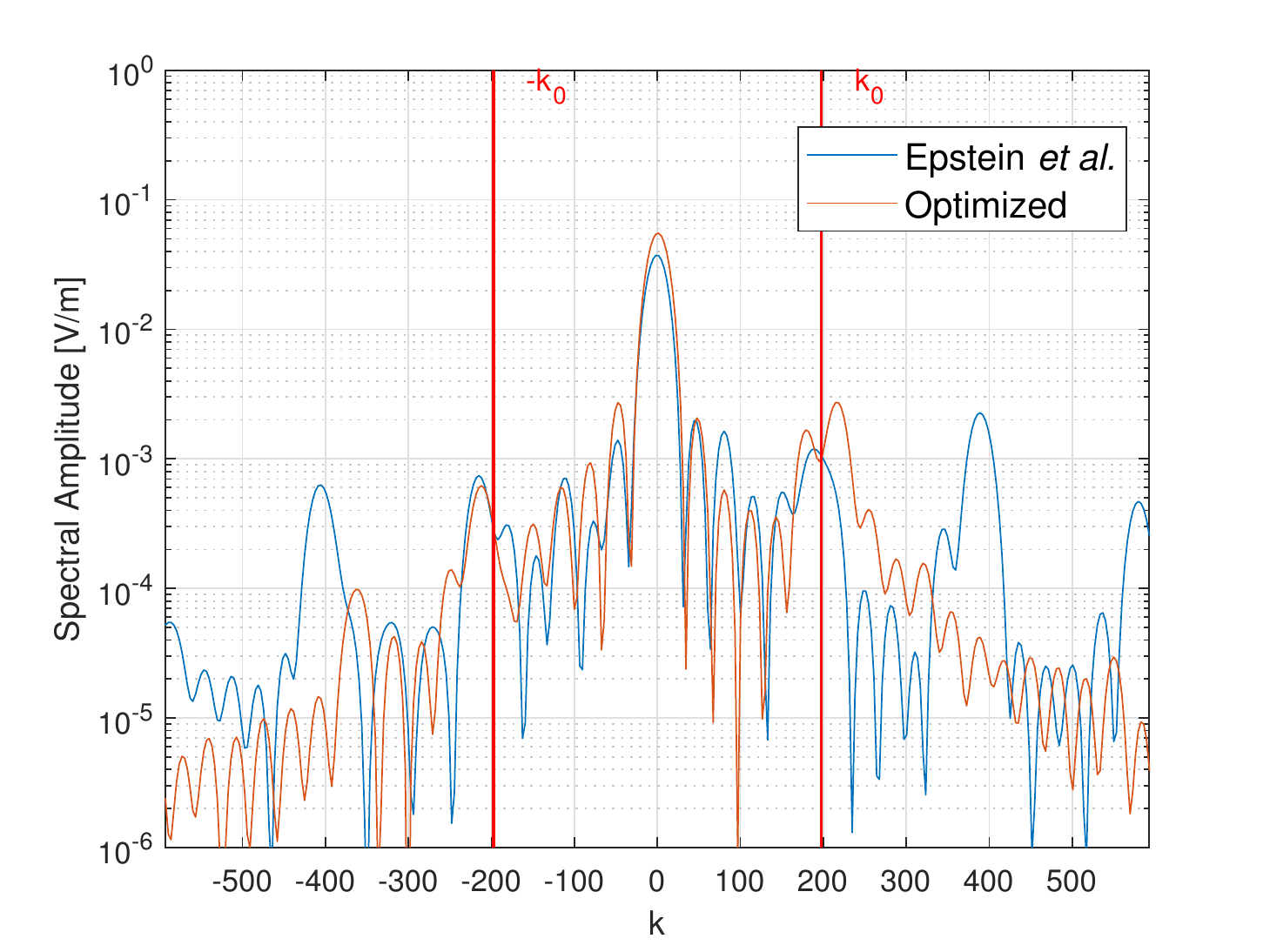}
         \caption{Electric field spectrum for $z > 0$}
         \label{fig:Espec_zpos}
     \end{subfigure}
     \hfill \\
     \begin{subfigure}[b]{0.45\textwidth}
         \centering
         \includegraphics[width=\textwidth]{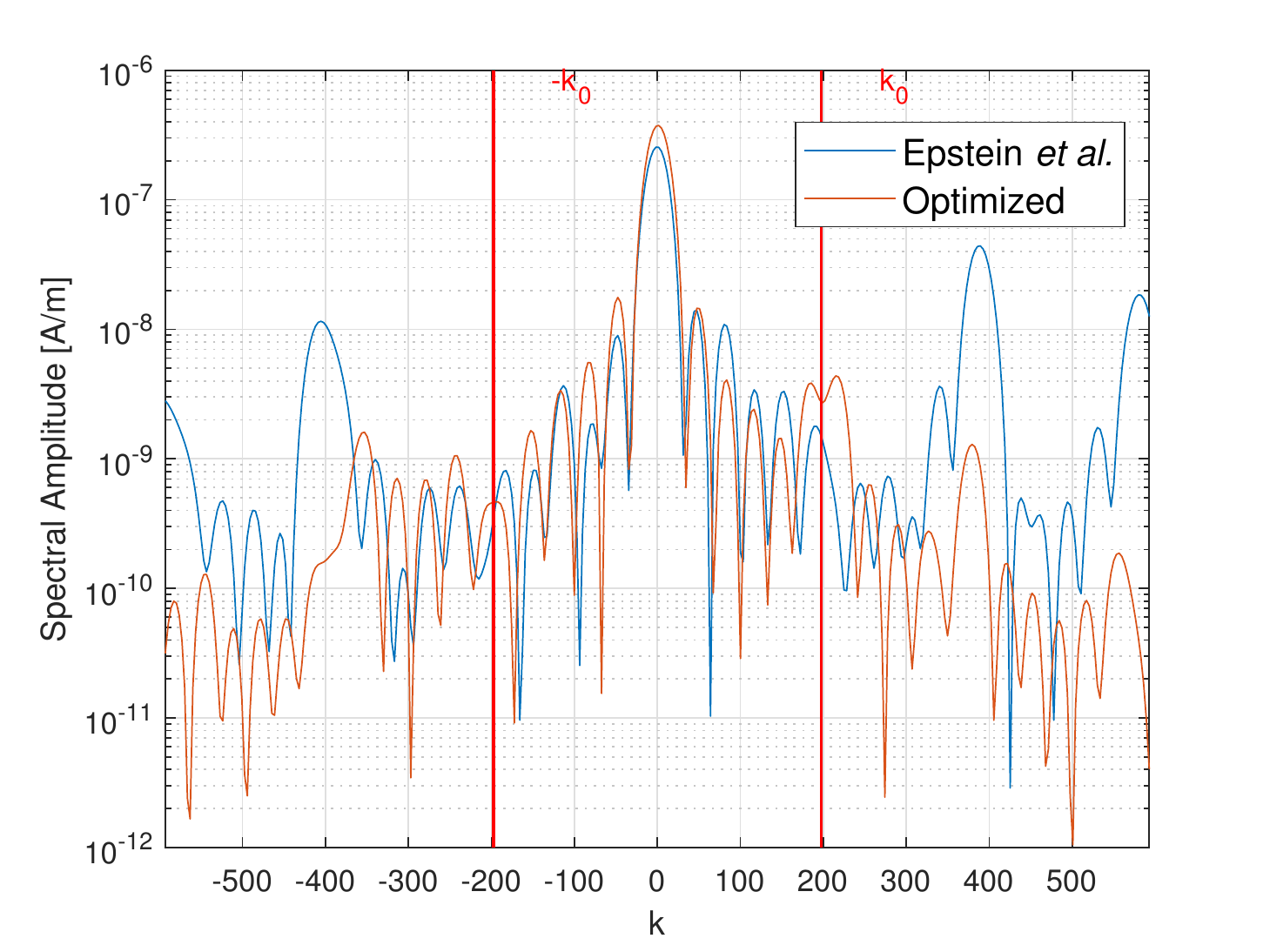}
         \caption{Magnetic field spectrum for $z > 0$}
         \label{fig:Hspec_zpos}
     \end{subfigure}
        \caption{Electric (a) and magnetic (b) field spectra for $z > 0$ (on the transmitted side of the EMMS)}
    \label{fig:EH_sepectrums_zpos}
\end{figure}

\begin{figure}
     \centering
     \begin{subfigure}[b]{0.45\textwidth}
         \centering
         \includegraphics[width=\textwidth]{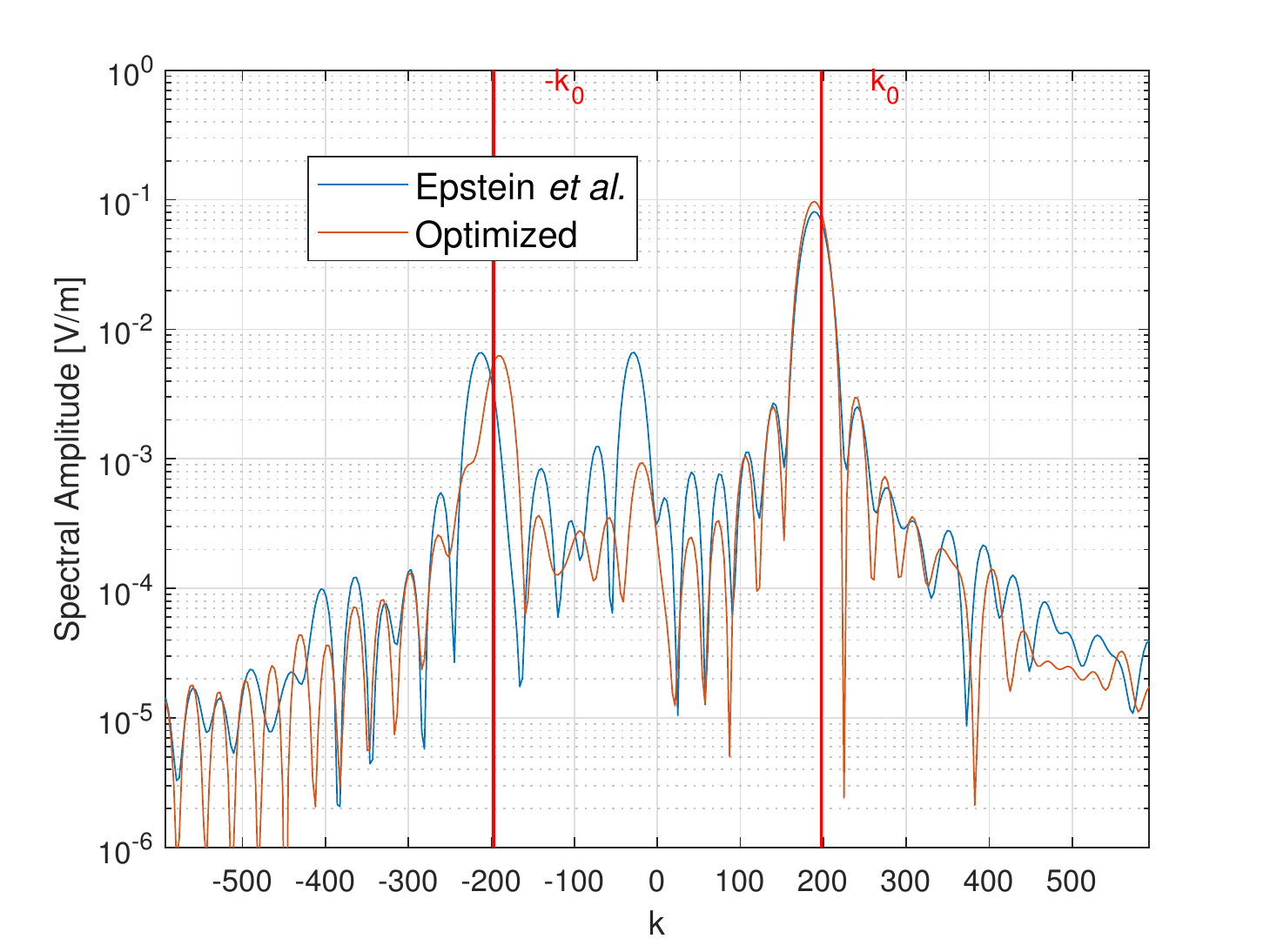}
         \caption{Electric field spectrum for $z < 0$}
         \label{fig:Espec_zneg}
     \end{subfigure}
     \hfill \\
     \begin{subfigure}[b]{0.45\textwidth}
         \centering
         \includegraphics[width=\textwidth]{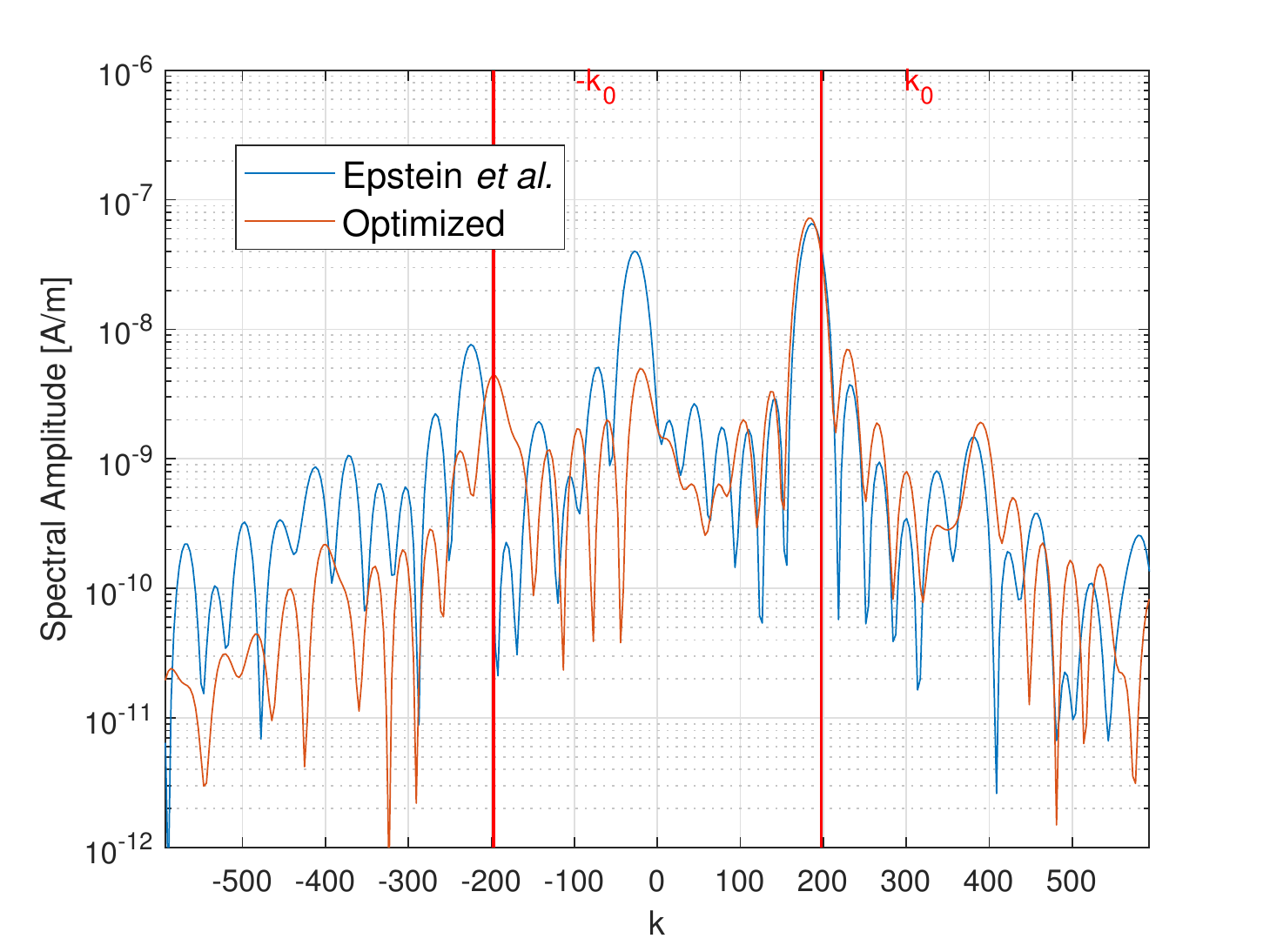}
         \caption{Magnetic field spectrum for $z < 0$}
         \label{fig:Hspec_zneg}
     \end{subfigure}
        \caption{Electric and magnetic field spectra for $z < 0$ (on the reflected side of the EMMS)}
    \label{fig:EH_sepectrums_zneg}
\end{figure}

In order to verify the extent to which AFs are leveraged, we examine the spectrum of the electric and magnetic fields just above and below the surface. We compare the analytic spectrum with the optimized one on top of and below the EMMS in \Cref{fig:EH_sepectrums_zpos} and \Cref{fig:EH_sepectrums_zneg} respectively. As expected, the analytic formulation, which explicitly excites AFs, has the expected peak in spectrum around $k=\pm 2k_0$, in order to perform the transformation \cite{Epstein2016b}. The optimized solution does not have the same obvious peak, however there is still an increased level of non-radiating spectrum with $k>|k_0|$. The differences are attributed to the finite nature of the EMMS, which requires different AFs to satisfy the problem goals than an infinite one.

% \begin{figure}[!t]
% \centering
% \includegraphics[width=0.5\textwidth]{images/Retro_Directivity.eps}
% \caption{Directivity of extreme angle reflection towards $\theta=108^\circ$ from an incident uniform illumination at $\theta = 0^\circ$. The optimized results are from the ADMM-based optimizer and analytical results are from Epstein \textit{et al} \cite{Epstein2016b} using AFs.}
% \label{fig:XTremeAngle_Directivity}
% \end{figure}

\begin{figure}[!t]
\centering
\includegraphics[width=0.5\textwidth]{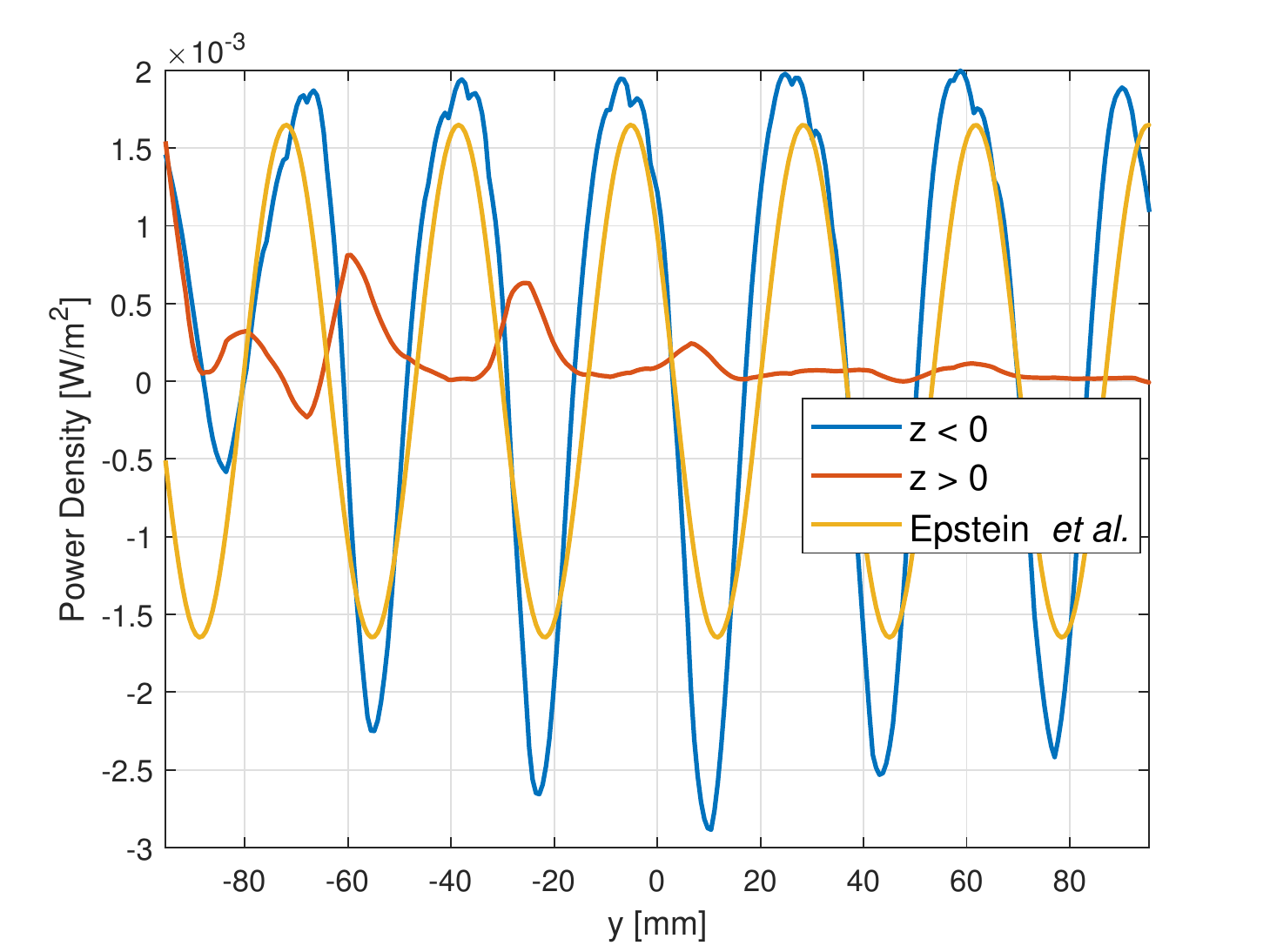}
\caption{Time averaged power density above and below the EMMS. The analytical results are from Epstein \textit{et al.} using AFs and arbitrary phase shift $\xi_{out} = 288^\circ$. \cite{Epstein2016b}}
\label{fig:PowerProfiles}
\end{figure}

We can further investigate the existence of AFs by examining the time-averaged power density redistributed above and below the EMMS. To do this, we calculate the near fields from our optimized surface currents in COMSOL. The time averaged power density can be seen in \Cref{fig:PowerProfiles} above and below the EMMS along with the analytically-predicted power profile by Epstein \emph{et al}. It is clear from analyzing the time-averaged power density at the input side of the surface that there is seemingly active/lossy behaviour without explicitly invoking the use of AFs with our passive/lossless EMMS. Interestingly, the spatial frequency of the power density profile matches relatively well between the input side of the EMMS and the analytically predicted result. Due to this example not achieving perfect reflection, there is naturally some time-averaged power density on the transmitted side of the EMMS ($z>0$).

\begin{figure}[!t]
\centering
\includegraphics[width=0.5\textwidth]{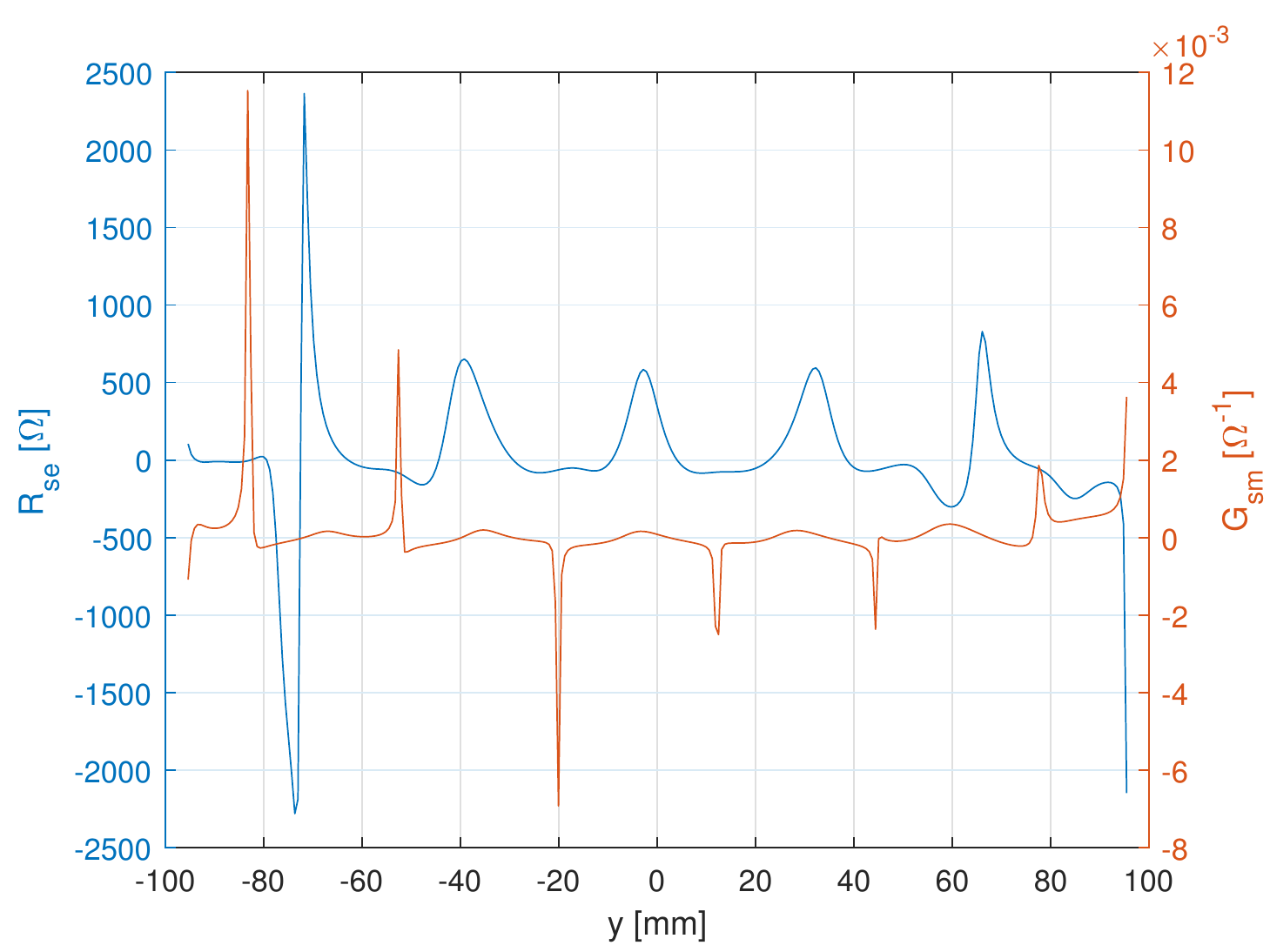}
\caption{Active and lossy components of $Z_{se}$ and $Y_{sm}$ ($R_{se}$ and $G_{sm}$ respectively) required to satisfy the MoM equality constraints for the extreme reflection example if the non-radiating portion of the surface current densities are removed.}
\label{fig:ZY_AL_refl}
\end{figure}

Lastly, we can quantify the extent that AFs are used if we isolate the non-radiating currents and remove them from the MoM matrix equations. Satisfaction of these MoM matrix equations is critical for maintaining a passive and lossless solution per their construction. Removing the non-radiating surface currents requires an active and lossy EMMS because the MoM equality constraints, \eqref{E_mom_eqn} and \eqref{H_mom_eqn}, are no longer satisfied for these new, purely radiating surface current densities. For this extreme reflection example, the non-negligible active and lossy components of the surface electric impedance and magnetic admittance are shown in \Cref{fig:ZY_AL_refl}. The two curves clearly show non-negligible real parts in the surface impedances and admittances. The bianisotropic coupling becomes a redundant degree of freedom with an active and lossy EMMS, so it is not shown. As expected, the non-radiating spectrum of the surface current densities chosen by the optimizer play a vital role in performing passive and lossless field transformations.

\section{MICROSCOPIC OPTIMIZER USING DNN SURROGATE MODELS}\label{sec:MicroOverview}
% details on surrogate models and how we train things (1-2 pages)
In the microscopic design step, the optimized set of $(Z_{se},Y_{sm},K_{em})$ are first converted to scattering parameters and then are realized using physical unit cells. To find the optimized unit cell, a previously proposed approach is employed \cite{Naseri2021a}.  

\begin{figure}[!t]
    \centering
    \includegraphics[width=\columnwidth]{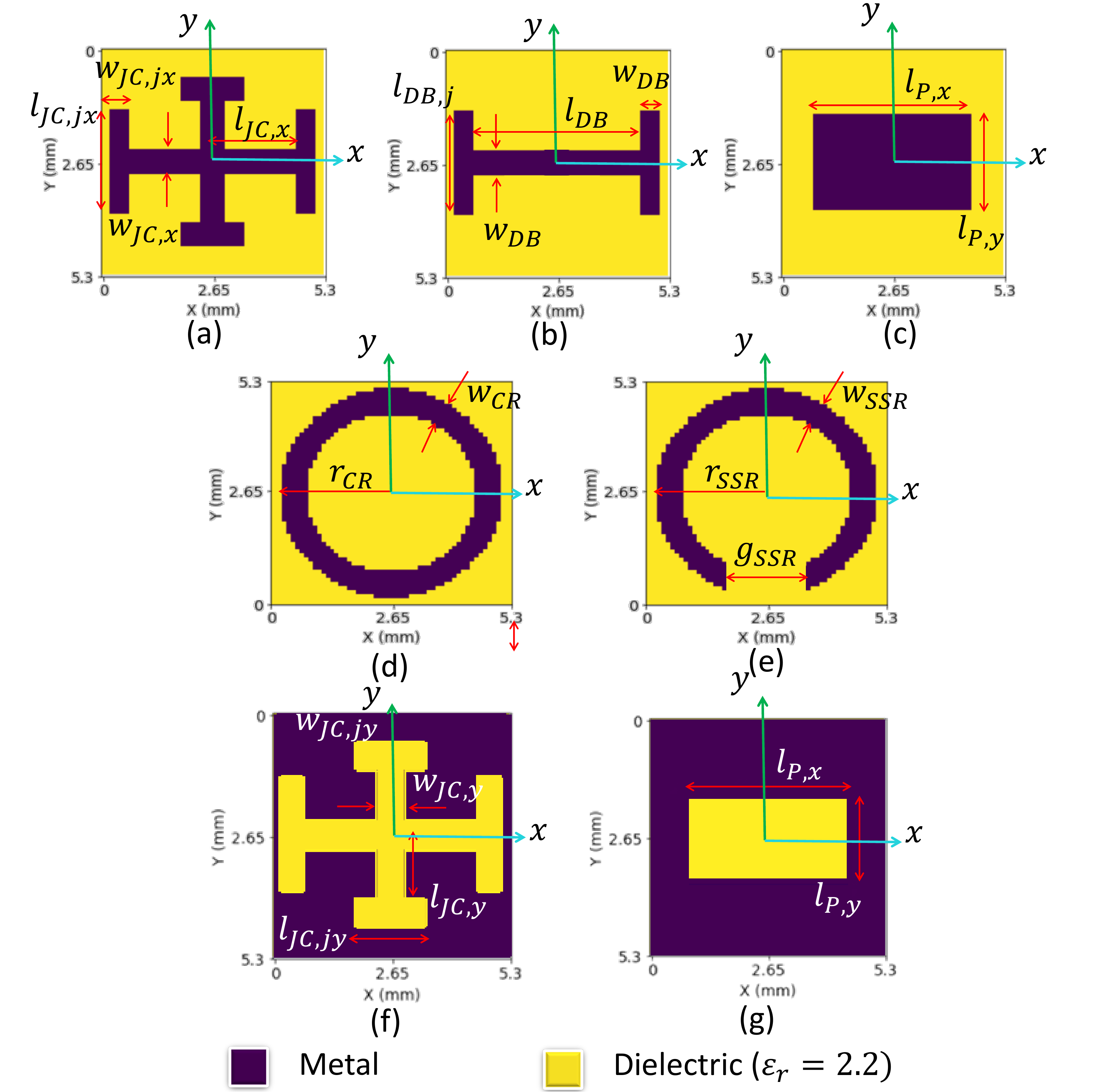}
    \caption{Primitives used to generate three-layer bianisotopic unit cells: (a)-(e) possible capacitive primitives on all the layers, and (f)-(g) additional inductive primitives on the middle layer. }
    \label{fig:primitives}
\end{figure}

First, about $70,000$ three-layer bianisotropic unit cells with a period of $5.3$ mm and composed of the primitives shown in \Cref{fig:primitives} are simulated using an accelerated in-house periodic MoM tool between $1.0$ and $19.0$ GHz. The scatterers on the top and bottom layers can be any of the primitives in \Cref{fig:primitives} (a)-(e) while the scatterer on the middle layer can be any of the primitives in \Cref{fig:primitives} (a)-(g). The scatterers are separated using RT Duroid $5880$ $(\epsilon_r=2.2)$ with an standard thickness in $\{0.254,0.508,0.787,1.524\}$ mm. Once these unit cells are simulated, their scattering parameters (S-parameters) for normal incidence of transverse electric (TE) and transverse magnetic (TM) fields are stored.

To predict the S-parameters of the unit cells composed of new combinations of the shown primitives, we train two  deep-learning neural networks (DNNs) comprising fully-connected layers of neurons as surrogate models. These networks are specifically adept at extracting the underlying features in large data sets and predicting their properties accurately. This is particularly useful in the case of thin EMMSs, where the inter-layer coupling between the scatterers of each unit cell is non-negligible. Therefore, in thin unit cells, interpolation of properties based on dimensions of the scatterers will likely cause inaccuracy in the predicted S-parameters.  

The two DNNs, \textit{mag}-DNN and \textit{phase}-DNN, are used to predict the magnitude and phase of the S-parameters, respectively. Both of them accept the frequency of interest and a \textit{feature} variable as the inputs. Here, the \textit{feature} variable represents the physical structure of the unit cell including the category of the constituent scatterers, (shown in \Cref{fig:primitives}), the dimensions of the scatterers, and the thickness of the substrate. The feature variable for each unit cell is composed of total of $30$ binary and continuous values. The binary values are used for the shape of the primitive and the standard thickness of the substrate, while the continuous values specify dimensions of features unique to each scatterer. The \textit{mag}-DNN simultaneously predicts the transmission and reflection coefficients of the TE and TM fields under normal incidence as four outputs. However, to predict the phase of each of these parameters, a separate DNN is trained for each. Moreover, each \textit{phase}-DNN receives one more additional input: the magnitude of the corresponding S-parameters at a certain frequency. Adding the magnitude of the S-parameter improves the accuracy of the DNN in predicting the phase response at the resonance frequencies. A \textit{phase}-DNN outputs three parameters that can be used to calculate the absolute phase of each unit cell \cite{Naseri2021a}.

\begin{figure*}
    \centering
    \includegraphics[width=0.9\textwidth]{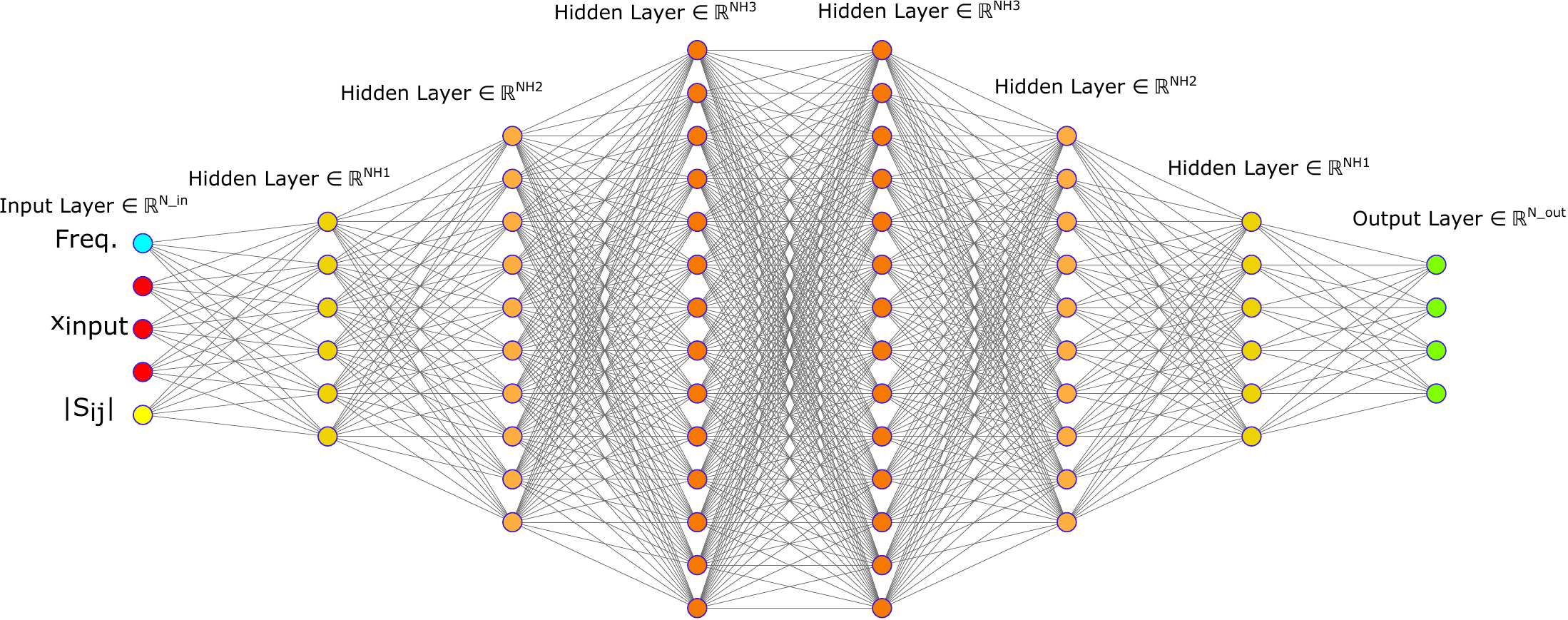}
    \caption{Representative architecture of the fully-connected neural networks used to predict the magnitude and phase of the scattering parameters of bianisotropic unit cells. The input node $|S_{ij}|$ is only connected in the \textit{phase}-NN.}
    \label{fig:neuralnetworks}
\end{figure*}

The \textit{mag}- and \textit{phase}-DNNs are composed of 6 hidden layers with $N_{H1}$, $N_{H2}$, $N_{H3}$, $N_{H3}$, $N_{H2}$, $N_{H1}$ neurons, respectively from the shallowest layer to the deepest layer of the neural networks, as shown in \Cref{fig:neuralnetworks}. The number of neurons for the \textit{mag}-DNN are $N_{H1}=100$, $N_{H2}=500$, and $N_{H3}=1000$ that end with $4$ output neurons with sigmoid activation function. The \textit{phase}-DNN is composed of more neurons, where $N_{H1}=100$, $N_{H2}=500$, and $N_{H3}=2000$. The neural networks are trained using the ADAM optimizer \cite{Diederik2015} and backpropagation method \cite{Rumelhart1986} on $85\%$ of the training set and tested on the remaining $15\%$. The training is performed for $100$ epochs with \textit{early stopping} set to become activated if no improvement in the accuracy is achieved after 10 epochs.

The two DNNs are then integrated in a particle swarm optimization (PSO) utilizing the surrogate models to evaluate the performance of a unit cell under test. For better convergence, the PSO is performed in the $30$-dimensional \textit{feature} space of the unit cells. At each iteration, for a certain particle in the swarm, the $30$-dimensional feature along with the frequency point of interest is first input to the \textit{mag}-DNN to predict the amplitude of the S-parameters, $|S_{ij}|$. Then, the frequency point, the $30$-dimensional feature, and the predicted $|S_{ij}|$, is fed to the \textit{phase}-DNN to predict the phase of the S-parameters. These predicted S-parameters are then compared to the required ones obtained from the macro-optimization design step to see if they meet the requirements. The process described  is repeated for the unique obtained sets of the $(Z_{se},Y_{sm}, K_{em})$ from the macroscopic design step to realize the whole EMMS with corresponding physical unit cells.

\section{DESIGN EXAMPLE: A BEAM-SPLITTING EMMS} \label{Sec:Example}

\begin{figure}
    \centering
    \includegraphics[width=\columnwidth]{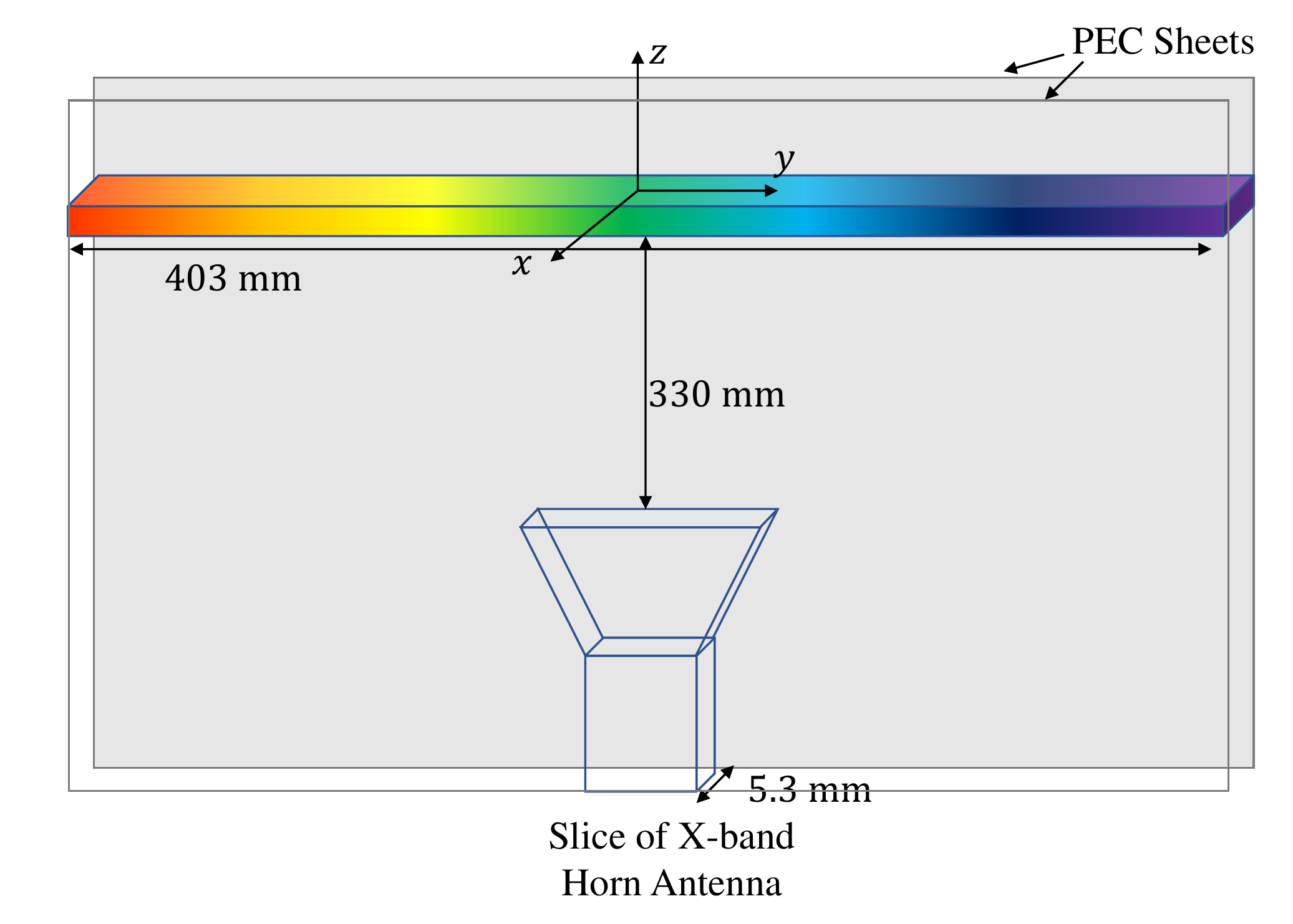}
    \caption{Simulation setup of the EMMS that is uniform in $x$-direction. The EMMS is uniformly excited by a slice of the horn antenna. Both are placed in a parallel plate waveguide between two perfect electric conductors. }
    \label{fig:2Dsimulation}
\end{figure}

Following the design process described, a one-dimensional $12.67\lambda_0$ ($403$ mm)-long metasurface varying along the $y$-direction is designed to radiate two beams at $\theta=-20^\circ$ and $\theta=+30^\circ$. The EMMS is uniform along the $x$-direction and the beam collimation occurs only in the $yz$-plane. This surface is composed of $76$ unique meta-atoms with a periodicity of $5.3 \;  \textrm{mm}=\lambda/6$, where $\lambda=31.8 \; \textrm{mm}$ at $9.4\; \textrm{GHz}$. Using the incident field produced by a realistic feed, a horn antenna, the optimization problem constraints are listed in \Cref{table:MultiCrit}. The feed antenna is modelled after a 3D X-band horn antenna discussed later by slicing it in the H-plane to mimic what will happen in the 3D experiment. The feed is placed $330 \:\textrm{mm}$ away from the center of the EMMS. In this configuration, the maximum angle of incidence at the edge of the EMMS is $31.4^\circ$. The simulation setup of the uniform EMMS being excited by a slice of the horn antenna is shown in \Cref{fig:2Dsimulation}.

\begin{table}[!t]
\caption{Optimization parameters for the two-beam EMMS example}
\label{table:MultiCrit}
\centering
\begin{tabular}{|c||c|}
\hline
 & \bfseries Value \\
\hline
\bfseries Wavelength ($\lambda$) & $31.8\; \textrm{mm}$ \\
\hline
\bfseries Surface Width ($W$) & $12.67 \lambda$ \\
\hline
\bfseries Incident Field  & 15 dBi Horn \\
\bfseries ($E^{inc}$) & Antenna\\
\hline
\bfseries Angular Sampling Points ($M$) & $361$ \\
\hline
\bfseries Spatial Sampling Points ($N$) & $76$ \\
\hline
\bfseries Max Iterations & $150$ \\
\hline
\bfseries Initial & $ ([\widetilde{\textbf{X}}_{se}]^0,[\widetilde{\textbf{B}}_{sm}]^0,[\widetilde{\textbf{K}}_{em}]^0) = \textbf{0}, $\\
\bfseries Conditions & $\rho = 200, \beta = 377, (\bm{\mu}_{\textbf{Z}_e}^0, \bm{\mu}_{\textbf{Z}_m}^0) = \textbf{0}$ \\
\hline
\bfseries $\left\lbrace \alpha_{MB},\alpha_{NU},\alpha_{D^E},\alpha_{D^H}\right\rbrace$ & $\left\lbrace 500,500,400,1\right\rbrace$ \\
\hline
\bfseries Main Lobe Angles ($MB$) & $\theta = \left\lbrace -20^{\circ}, 30^{\circ} \right\rbrace$ \\
\hline
\bfseries Main Lobe Level ($MB_{level}$) & 2.65 [V/m] \\
\hline
 Sidelobe  & $ \lbrace  -12 \textrm{dB}, -90^{\circ}\leq \theta\leq -28^{\circ}\rbrace ,$\\
 Level ($\bm{\tau}$)& $ \lbrace  -12 \textrm{dB},  -12^{\circ}\leq \theta\leq 21^{\circ}\rbrace ,$ \\
 and Angles ($SL$)  &  $ \lbrace -12 \textrm{dB},  39^{\circ}\leq \theta\leq 90^{\circ}\rbrace , $\\
 & $  \lbrace -20 \textrm{dB},  90^{\circ}\leq \theta\leq 180^{\circ}\rbrace ,$\\
 &  $ \lbrace -20 \textrm{dB},  -180^{\circ}\leq \theta\leq -90^{\circ}\rbrace ,$\\
\hline
 Null Angles ($NU$)& $\theta=\left\lbrace 0^{\circ},180^{\circ}\right\rbrace$  \\
\hline
\bfseries $\left\lbrace \textbf{D}^e_{max},\textbf{D}^m_{max} \right\rbrace$ & $\left\lbrace \textbf{0.01},\textbf{1}\right\rbrace$ \\
\hline
\end{tabular}
\end{table}

\begin{figure}
    \centering
    \includegraphics[width=\columnwidth]{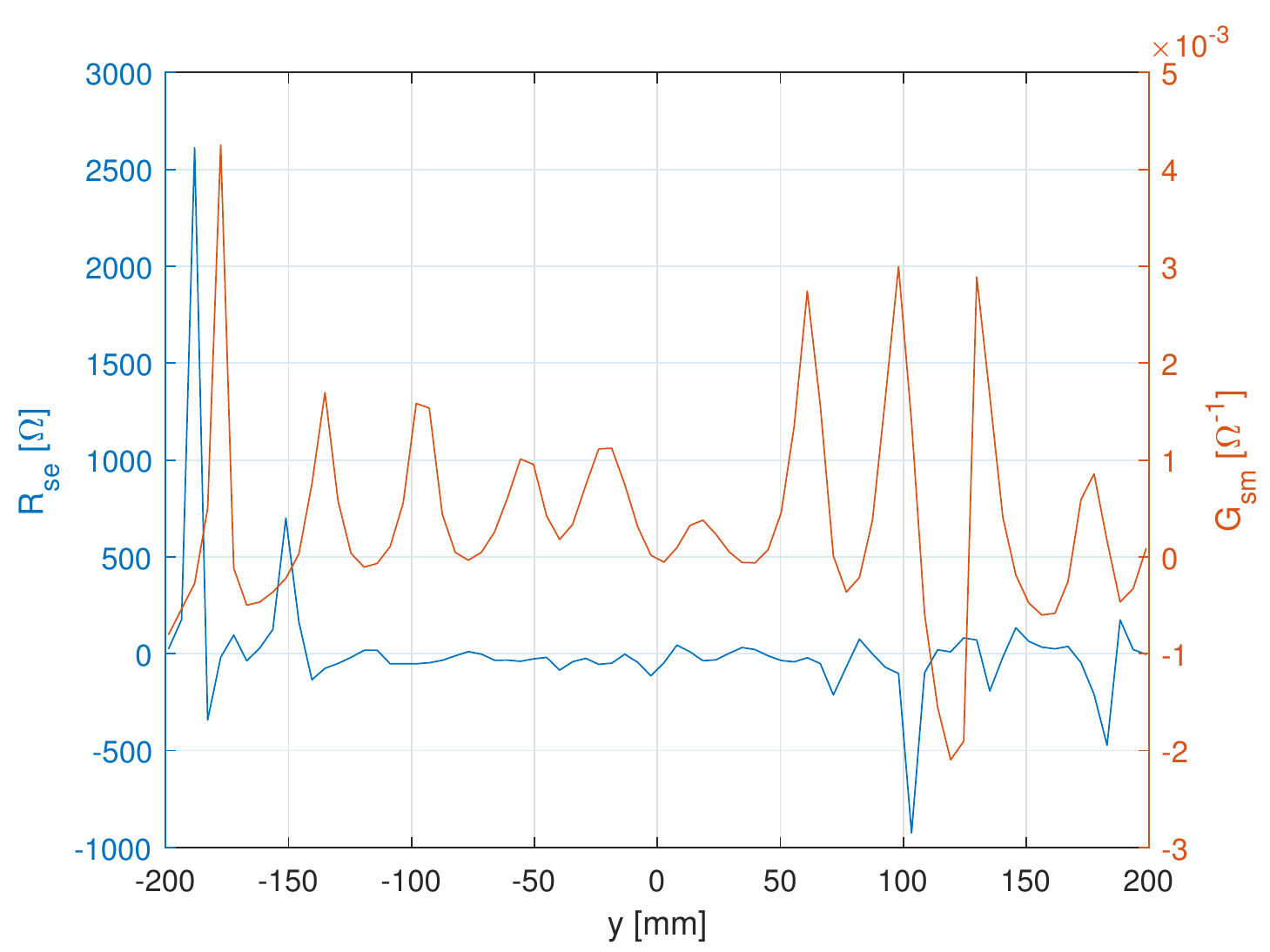}
    \caption{Required active and lossy surface parameters if AFs are removed for the two-beam EMMS example.}
    \label{fig:ZY_AL_exp}
\end{figure}

It is worth noting that the AFs are essential for this optimization problem, because if we remove the non-radiating currents from the homogenized model, the EMMS is required to be active and lossy with real surface electric impedance and magnetic admittance. These active and lossy surface parameters are pictured in \Cref{fig:ZY_AL_exp}. 

\begin{figure}
    \centering
    \includegraphics[width=\columnwidth]{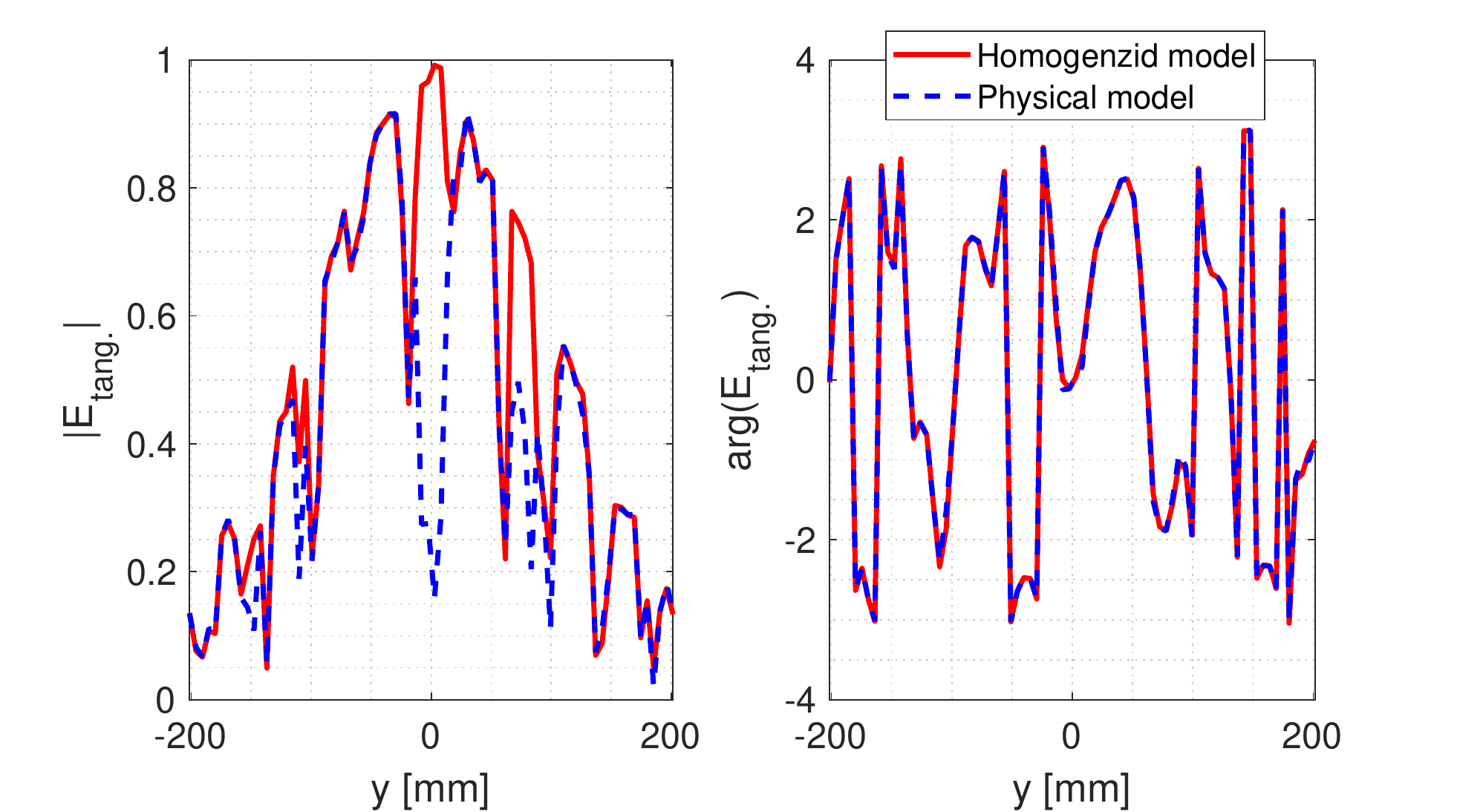}
    \caption{Comparison between the amplitude and phase of the tangential components of the transmitted electric field from the EMMS with the optimized macroscopic and microscopic surface properties at $9.4$ GHz. }
    \label{fig:tanEy}
\end{figure}

The comparison between the amplitude and phase of the tangential electric field component on the transmitting side of the optimized homogenized model and the physical EMMS is shown in \Cref{fig:tanEy}. There is good agreement between the desired phases of the homogenized model and the physical unit cells. The amplitude is also well matched for the most part, except at two points. Matching the desired amplitude at $y=0$ and at $y=77.9\; \textrm{mm}$ requires realizing completely inductive surface impedances, i.e. $\Im\{Z_{se}\}>0$ while $\Im\{Y_{sm}\}=\Re\{K_{em}\}=0$, by the meta-atom. Since the possible generated meta-atoms in the macroscopic design step have complementary-shaped scatterers in the middle layer but not the top and bottom layers, this may have made it difficult to achieve the required impedances here. 

% The far field patterns of the S-parameters and the homogenized model are shown in \Cref{fig:FF_Compare_Experiment} and compare favourably. From this we can be confident that matching the desired S-parameters will give us a pattern close to that which we originally designed for.

% \begin{figure}
%     \centering
%     \includegraphics[width=\columnwidth]{images/FF_Compare_Experiment_Trans.eps}
%     \caption{Comparison between the far field patterns generated from the S-parameters and homogenized model. }
%     \label{fig:FF_Compare_Experiment}
% \end{figure}

% TODO: Show/quantify the extent to which AFs are used to satisfy this design. This could be in the form of
% the AF spectrum or just saying how much the equality constraints are violated without AFs

\section{EXPERIMENTAL VALIDATION OF SYNTHESIS METHOD} \label{sec:ExperimentalVerif}
% Show the experimental setup from the optimization stage all the way to our near field chamber configuration. Present our impeccable results to show how great we are (1-2 pages)

\begin{figure}
    \centering
    \includegraphics[width=0.9\columnwidth]{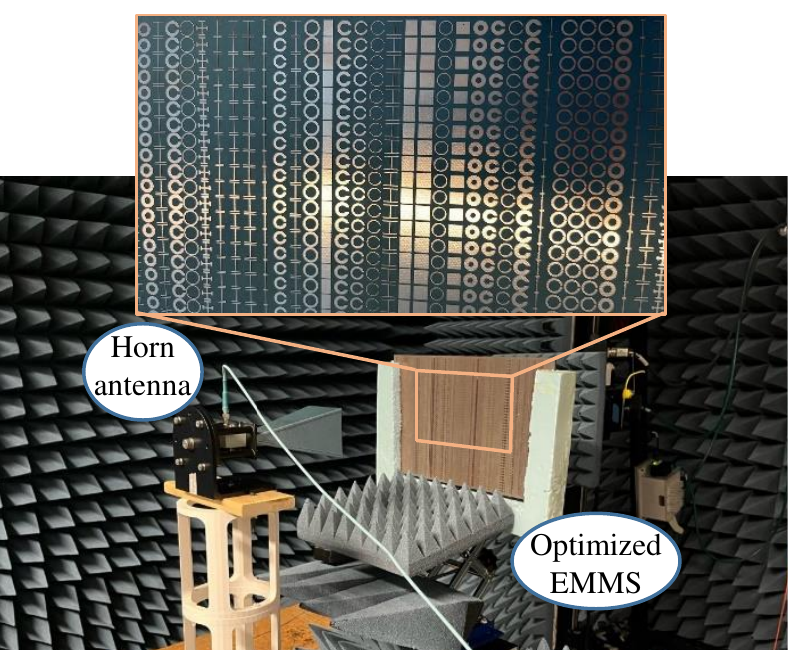}
    \caption{The prototype in the measurement setup. }
    \label{fig:prototype_setup}
\end{figure}

A prototype of the optimized bianisotropic metasurface with the width of $402.8$ mm and length of $243.8$ mm is fabricated and tested in a planar near-field antenna scanner using the setup shown in \Cref{fig:prototype_setup}. Like the simulation model, this surface has $76$ unique meta-atoms varying along the width in the $y$-direction but is uniform in the $x$-direction. There are $46$ unit cells in the $x$-direction. A detailed picture of the prototype in this setup is shown in the inset of \Cref{fig:prototype_setup}, where it can be seen that the designed EMMS is composed of meta-atoms with many different scatterer shapes. 

\begin{figure}
    \centering
    \includegraphics[width=0.9\columnwidth]{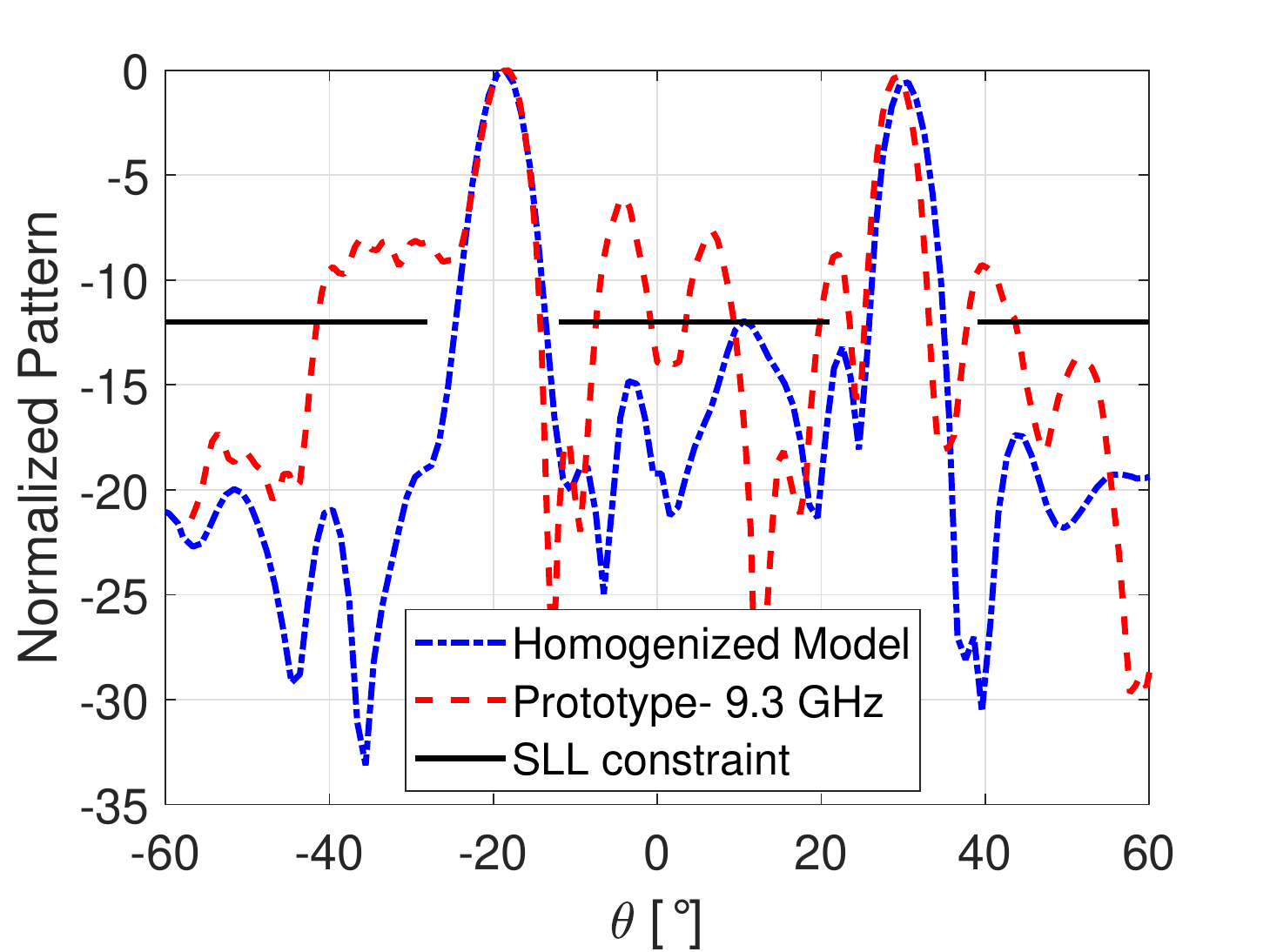}
    \caption{Normalized pattern of the homogenized model at 9.4 GHz and the prototype at 9.3 GHz in $xz$-plane $(\varphi=0^\circ)$ plane.  }
    \label{fig:FF_pattern}
\end{figure}

The normalized radiation patterns of the homogenized model and the prototype are shown in \Cref{fig:FF_pattern}, where both patterns have two beams directed at $\theta=-20^\circ$ and $\theta=+30^\circ$. The SLL of the homogenized model's pattern complies with the specified maximum level of $-12 \; \textrm{dB}$, whereas the prototype's SLLs are slightly higher than $-10 \:\textrm{dB}$. The higher SLLs in the prototype can be attributed to two factors. The first one is the difference between the optimized macroscopic properties and the ones realized by the physical meta-atoms. This occurs when the required surface properties cannot be completely realized by one of the meta-atoms in the considered three-layer solution space. For the example here, one could explore an augmented space that includes inductive scatterers similar to \mbox{\Cref{fig:primitives}} (f)-(g) on all three layers. The second is that although mutual coupling is accounted for in the macroscopic optimization stage, the unit cells for the microscopic optimization stage are optimized for in a periodic environment. This assumption is valid if the unit cells are not changing drastically spatially, which there is not guarantee of in this case. It is worth noting that since the maximum angle of incidence on the EMMS is $31.4^\circ$, the performance of the constituent meta-atoms under oblique incidence including and up to $31.4^\circ$ is stable and close to its properties under normal incidence.

\begin{figure}
    \centering
    \includegraphics[width=0.9\columnwidth]{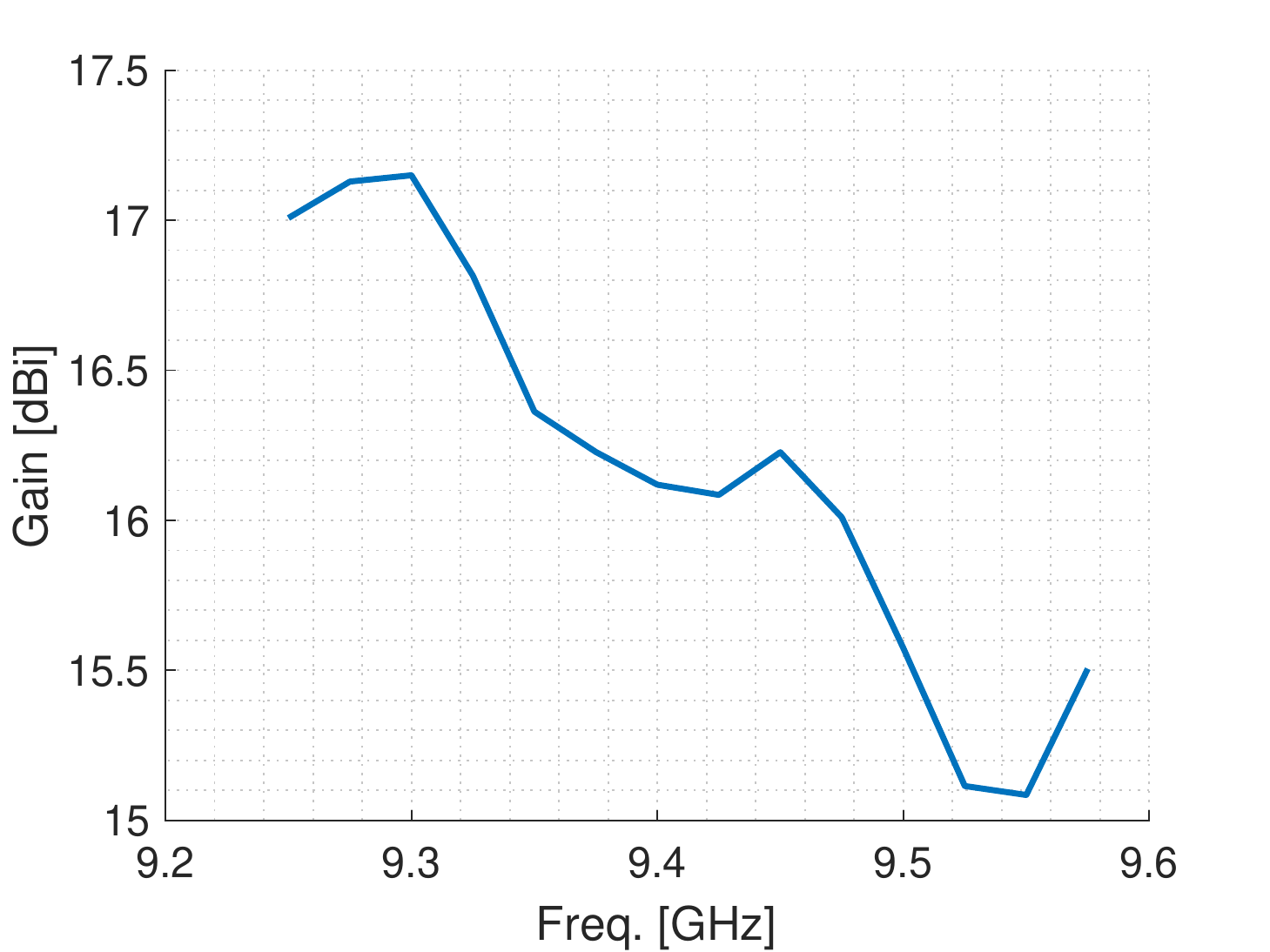}
    \caption{Gain of the fabricated EMMS at $\theta=-20^\circ$ and $\varphi=0^\circ$.  }
    \label{fig:gain}
\end{figure}

In order to quantify the gain of the surface, we compared the field levels to that of a standard X-band gain horn over the frequencies from $9.2$ GHz to $9.5$ GHz. The gain of the horn antenna with the prototype in front of it at $\theta=-20^\circ$ is shown in \Cref{fig:gain} as a function of frequency. The maximum gain is found to be $17.2$ dBi at $9.3$ GHz. In order to determine the merit of this gain figure compared to the theoretical maximum, we itemize the factors that reduce the realized gain at $\theta=-20^\circ$ in \Cref{table:loss}. 

% More EMMS loss notes: 
% * Should I center the Horn NF data by amplitude or just use it as is? I'm using the data we gathered right after removing the MS I think so the horn is a bit skewed: Yes, I think you should use clean data so center it. Was the horn 330 mm away from the probe? Otherwise projected data should be used 

% Collimation only in azimuth: Calculate directivity from MS with uniform amplitude but phase from horn varying in elevation plane. Uniform phase in azimuth. So phase copied in x direction but varies in y.: Yes, that's perfectly correct 

% Taper loss: calculate directivity for MS with above phase variation and amplitude taper from horn in x-direction copied. Amplitude for each y wouldn't change.: When calculating taper loss, usually only amplitude is considered. So, phase would be a constant number (you can set it to zero everywhere). The same for the reference aperture (constant amplitude and phase everywhere)

% Spillover loss: calculate total far field power difference from total measured horn vs. horn across the extent of the EMMS: yes, the scannning window is larger than the MS and it should include all the power from the horn. So please note that the NF amplitude is tapered more than 12 dB  from both sides in the scanning window 

An ideal uniformly illuminated surface with the same length and width of our prototype radiates a pencil beam with a directivity of $30.7$ dBi. In our design, this is reduced $5.8$ dB by the fact that the EMMS only collimates in the $xy$-plane. Furthermore, the taper and spillover loss for the planar rectangular EMMS are calculated \cite{2013ReflectarrayAI} and amount to $1.4$ dB. In addition, since the beam is being split into two directions, the gain is reduced by approximately $3.0$ dB. Scanning the beams to $-20^\circ$ and $+30^\circ$ results in minimum of $10\log_{10}[\cos(-20)^\circ]=0.3$ dB scanning loss. Furthermore, the HFSS model yields a transmission efficiency of $74.5\%$, which results in $1.3$ dB of transmission loss including the dielectric loss. This transmission efficiency is calculated by comparing the feed's power alone and the power of the feed plus the EMMS composed of perfect electric sheet scatterers on a line on top of the EMMS. Lastly, there is $0.7$ dB of ohmic losses from the synthesized meta-atoms comprising scatterers of $0.018\;\textrm{mm}$-thick copper. To calculate the ohmic losses, first, the meta-atoms composed of copper scatterers under oblique incidence based on their position were simulated with periodic boundary conditions and their scattering parameters were extracted and compared to the case where perfect conductors are used. All of these factors lead to the $18.2$ dBi of expected gain from the prototype. With these considerations, there is only $1.0$ dB of discrepancy between the predicted and actual measured gain, which is within an acceptable range. The discrepancy can be attributed to manufacturing differences between the HFSS model and the prototype. In addition, there are also some inevitable small phase errors introduced by inaccuracies in the measurement setup. 

\begin{table}[!t]
\caption{Loss Analysis of the Prototype at $\theta=-20^\circ$}
\label{table:loss}
\centering
\begin{tabular}{|c||c|}
\hline
 & \bfseries Value [dB] \\
\hline
\bfseries Directivity of uniform surface   & $30.7$ \\ % reference surface with uniform amplitude and phase
\bfseries with $W=12.67\lambda$ \& $L=7.83\lambda$ & \\
\hline
\bfseries Loss due to collimation only in the azimuth direction & $5.8 $ \\ % uniform amplitude but copy phase offset vertically from source. Basically constant phase in x-direction for each y. I'm only getting 2.5 dB for actual horn
\hline
\bfseries Taper loss & $0.9 $ \\ % directivity when you copy the amplitude taper vertically and horizontally from the horn but keep phase constant. Compare this with the totally uniform example. I got -0.9dB with the actual horn data as well
\hline
\bfseries Spillover loss & $0.5$ \\ % calculate total far field power from horn and then horn across MS only. I got -0.3 spillover (going to check again) with the actual horn
\hline
\bfseries Beam splitting & $3.0$ \\ % divide by 2
\hline
\bfseries Scanning loss & $0.3$ \\ %10*log10(cos(theta)), theta = 20 deg
% \hline
% \bfseries Homogenized model's transmission loss & $0.4$ \\ % 0.7826 = 10*log10(0.907)
% \hline
% \bfseries Mismatch between optimized & $0.5$\\
% \bfseries macro and micro properties &  \\
\hline
\bfseries Transmission Loss of HFSS Model with PEC traces & $1.3$ \\ %10*log10(0.745) = 1.3dB
\hline
\bfseries Ohmic losses & $0.7$ \\
\hline 
\hline
\bfseries Calculated gain & $18.2\: \textrm{dBi}$ \\
\hline
\bfseries Measured gain & $17.2\: \textrm{dBi}$ \\
\hline
\end{tabular}
\end{table}

\section{CONCLUSION} \label{sec:Conclusion}
% Tie all the stuff together 
In this paper we have both determined the extent to which our previously reported EMMS optimizer leverages AFs and experimentally verified our EMMS inverse design scheme. The use of AFs was analyzed by performing a challenging extreme-angle reflection to $108^\circ$. We then qualitatively evaluated the presence of AFs by examining the near field radiating and evanescent spectrum. To further quantify the existence of AFs, we used SVD to isolate the non-radiating spectrum and evaluate its importance in satisfying the passive and lossless GSTCs. Both methods reveal that AFs are vital as a degree of freedom for the macroscopic optimizer to perform the field transformations studied.

We then used this method to design an EMMS based on far-field goal in the form of two equal beams at $\theta = -20^\circ$ and $30^\circ$ with a side lobe level of $-12$ and $-20$ dB. These surface parameters were fed into the machine learning-based microscopic optimizer to design a set of physical meta-atoms. These meta-atoms were obtained without the need for heuristically tuning constituent scatterer geometries. This surface was then fabricated and evaluated in an experimental measurement. The prototype's far-field pattern matches well with the original main beam direction and side lobe specifications. 

This end-to-end methodology yields good results without the designer needing to iteratively tune meta-atom physical properties. Furthermore, we have been able to experimentally verify the results with good agreement with a theoretical maximum gain. However, there are some areas to improve upon this promising EMMS design method. Firstly, the macroscopic optimizer is still configured for a two-dimensional EMMS. Moving to three dimensions will allow for more sophisticated beam forming and collimation in the elevation plane. Secondly, adding a minimum gain mask to the macroscopic design step will allow for specification of more sophisticated design goals such as isoflux and cosecant patterns. Thirdly, although mutual coupling is captured in the homogenized macroscopic model, the meta-atoms in the microscopic model are still selected based on their scattering parameters with the local periodic assumption and normal incidence. Further refinement in this avenue will allow for more accurate capture of the mutual coupling between elements yielding better agreement between optimizers. Lastly, the optimized EMMS is quite narrow band as bandwidth was not explicitly optimized for from the beginning. Further work to optimize for a desired bandwidth would be valuable. 

\section*{ACKNOWLEDGMENT}
We would like to thank Prof. George Elefteriades and Vasileios Ataloglou for many extremely helpful discussions concerning auxiliary fields.

\bibliographystyle{IEEEtran}
\bibliography{IEEEabrv,library}
\end{document}